\documentclass[a4paper,11pt]{article}
\usepackage{jheppub} 
\usepackage{lineno}
\usepackage{amsmath,amsfonts}
\usepackage{indentfirst}
\usepackage{hyperref}

\title{\boldmath
Radiation-like Shock Waves in Kink Scattering}

\author[a,1]{Xiang Li\note{Corresponding author.}}
\author[b]{Lingxiao Long}

\affiliation[a]{ Department of Physics and Astronomy Stony Brook University Stony Brook, NY, 11794, USA}
\affiliation[b]{School of Space Science and Physics, Shandong University at Weihai, Weihai, 264209, Shandong, China}

\emailAdd{xiang.li.9@stonybrook.edu, llxiao@mail.sdu.edu.cn}

\abstract{We study the radiation in kink collision via a model that varies between $\phi^6$ theory and $\phi^2$ theory with some discontinuities. Both numerical and analytical methods were used to investigate The kink-antikink(KAK) and antikink-kink(AKK) collision. In the numerical analysis, we found the critical velocities in both collisions increased with $n$. We also found a finite lifetime oscillon window in KAK collision for $n=2$. In the analytical part, we found a family of shock wave solutions that describes radiation in the kink collision perfectly. Moreover, an analytical AKK solution at $n\rightarrow\infty$ and $v=1$ was found by considering a certain limit of these solutions. }

\begin{document}
\maketitle
\flushbottom

\section{Introduction}

  Kinks are localized solutions of $1+1$ dimensional classical field equations with finite total energy. They connect different vacua and cannot decay into radiation completely.  Kinks appear in various contexts, ranging from cosmology \cite{Vachaspati:2006zz,COS1}  to field theory\cite{F1,F2,F3} and condensed matter physics\cite{Chaikin_Lubensky_1995,CON2}. 
  
  The dynamics of multiple kinks and antikinks exhibit various and complicated behaviors.   A particularly interesting case is the scattering of a kink-antikink pair. In integrable models, like the Sine-Gordon theory, the kink and antikink pass through each other, acquiring a phase shift. However, in non-integrable models like $\phi^4$ and most polynomial potentials, the kink-antikink collision can be significantly more complicated. For instance, in $\phi^4$ kink-antikink collision, there exists a critical incident velocity $v_c$. Above $v_c$, the kink and anti-kink get reflected, while under $v_{c}$, they either form a bion or get reflected depending on different incident velocities. Moreover, the outgoing window, with incident velocity as the independent variable, forms a fractal structure. This intricate dynamic behavior is thought to be generated from the energy exchange between vibrational and translational modes. 

Kink scattering exhibits even richer phenomena in higher polynomial models. In polynomial models such as $\phi^6$, $\phi^8$, and $\phi^{10}$, besides the $\phi^4$-like behavior, they can also display sine-Gordon-like behavior, completely decayed bions behavior, and $\phi^4$-like behavior without fractal structures\cite{shortrange}\cite{Exo}. Among these models, the completely decayed solutions are worth noting. A notable feature is that the boundary always moves close to the speed of light, regardless of incident velocity. This phenomenon implies a mechanism allowing the massive scalar field to travel at the speed of light.

Inspired by these higher polynomial models, we propose a new model, the quadratic half-compact model, which generalizes the $\phi^6$ model. "Half-Compact" means one side of the kink solution reaches the vacuum at a finite value, while the other side reaches the vacuum at infinity. Since the collision shows a simpler behavior on the compact side, this model allows us to examine the vacuum oscillation on the non-compact side. Moreover, due to the additional integrability at $n\rightarrow \infty$, we find an analytical solution that describes the completely decayed solution perfectly. We further discover that even in $\phi^4$ and $\phi^6$ theories,  the radiation exhibits similar behaviors. 

 The outline of this paper is as follows: Section \ref{themodel}  introduces the quadratic half-compact model. In Section \ref{PDE}, we show the numerical result of the kink-antikink and antikink-kink collision.  Section \ref{Analytic} provides an analytical solution and a test of the similarity between the analytical and the numerical results. Finally, we conclude our work in Section \ref{conclusion}.
  
\section{The Quadratic Half-Compact Model}
\label{themodel}
\begin{figure}[htbp]
\centering
\includegraphics[width=.8\textwidth]{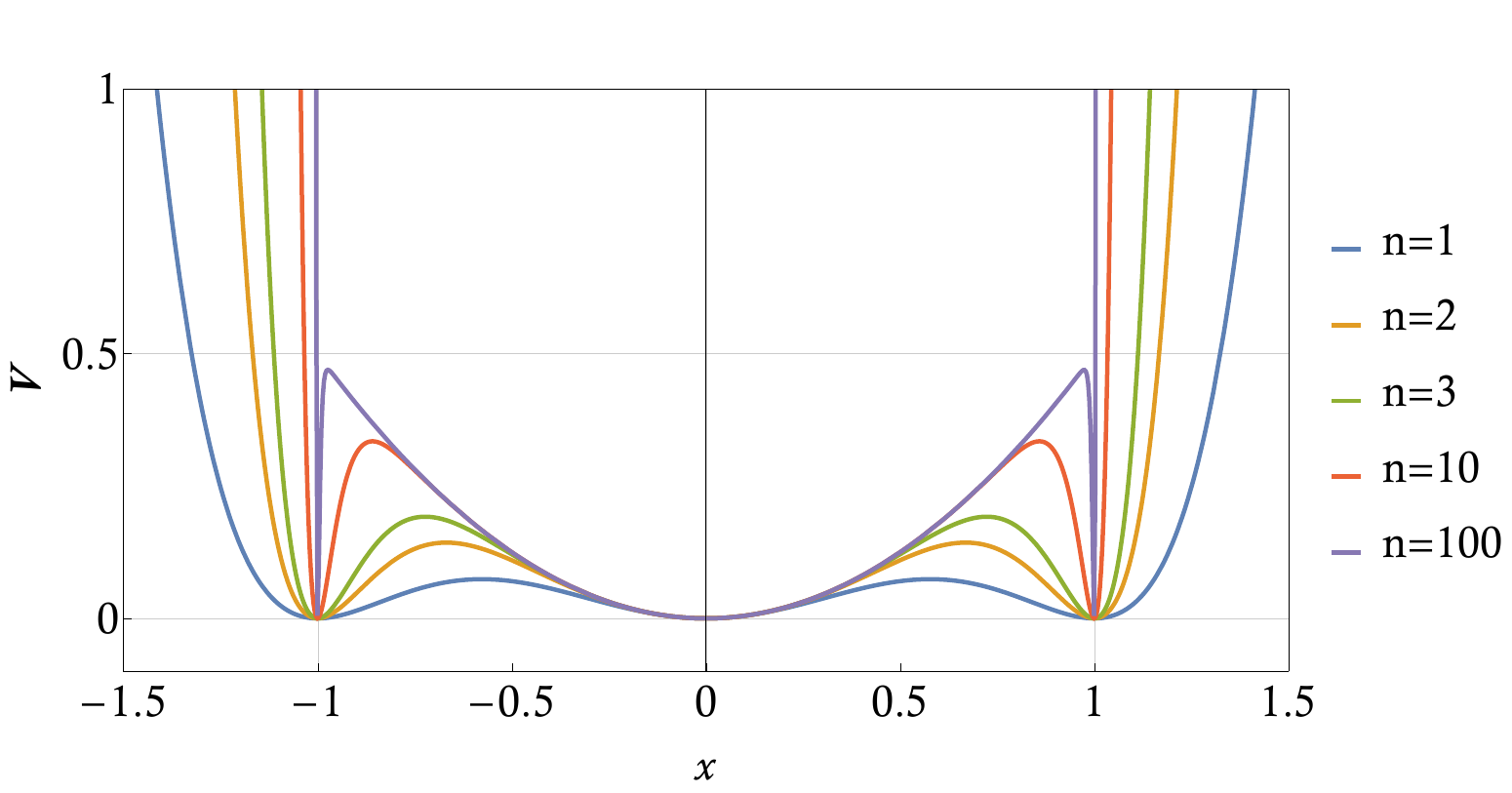}
\qquad
\caption{Examples of the potential for different $n$.  We set $k=1$ for this and all the figures below.
\label{U}}
\end{figure}

\begin{figure}[htbp]
\centering
\includegraphics[width=.8\textwidth]{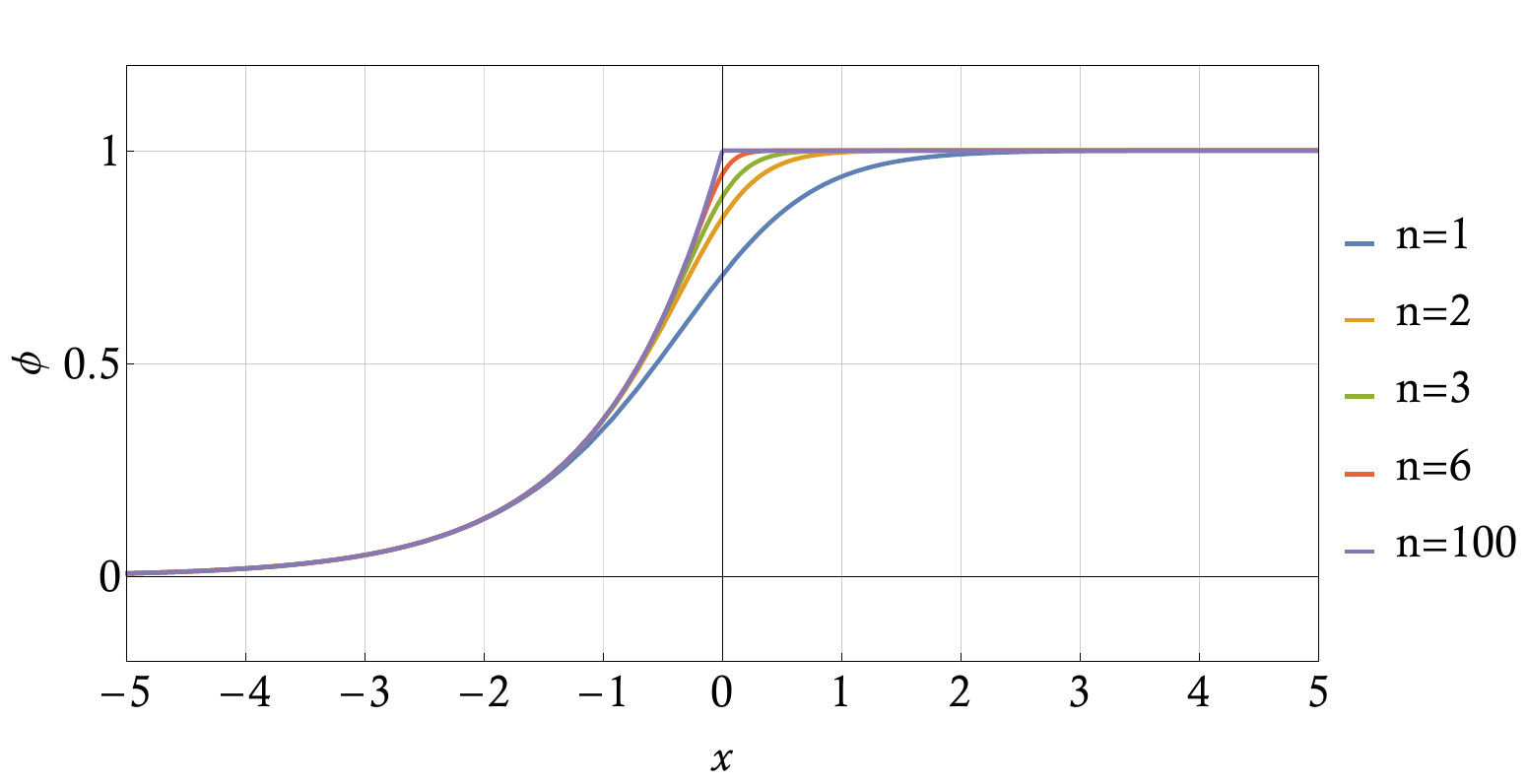}
\qquad
\includegraphics[width=.8\textwidth]{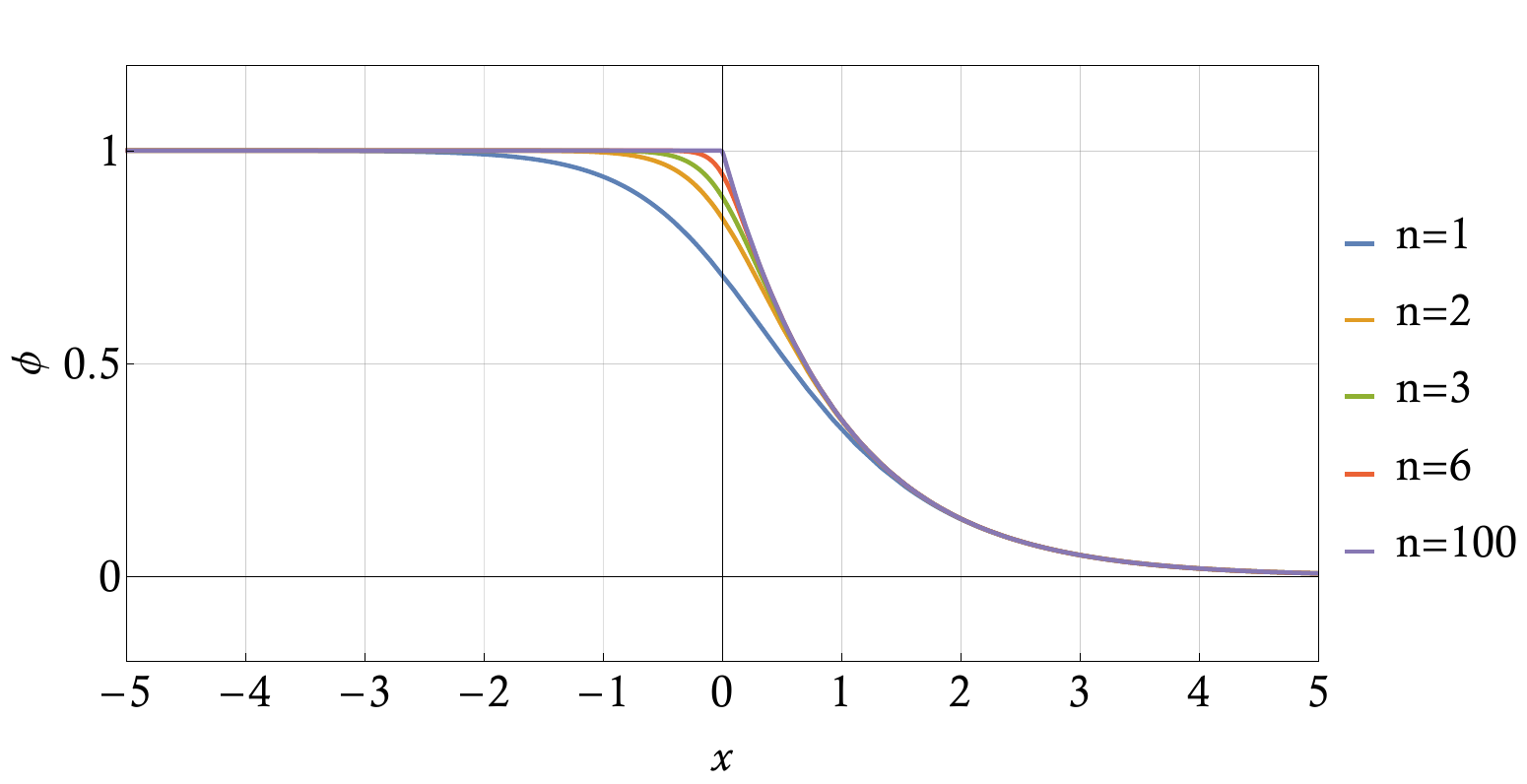}
\qquad
\caption{ The configurations of kink (top) and antikink (bottom) with $x_{0}=0$ for different $n$.
\label{KF}}
\end{figure}

\begin{figure}[htbp]
\centering
\includegraphics[width=.8\textwidth]{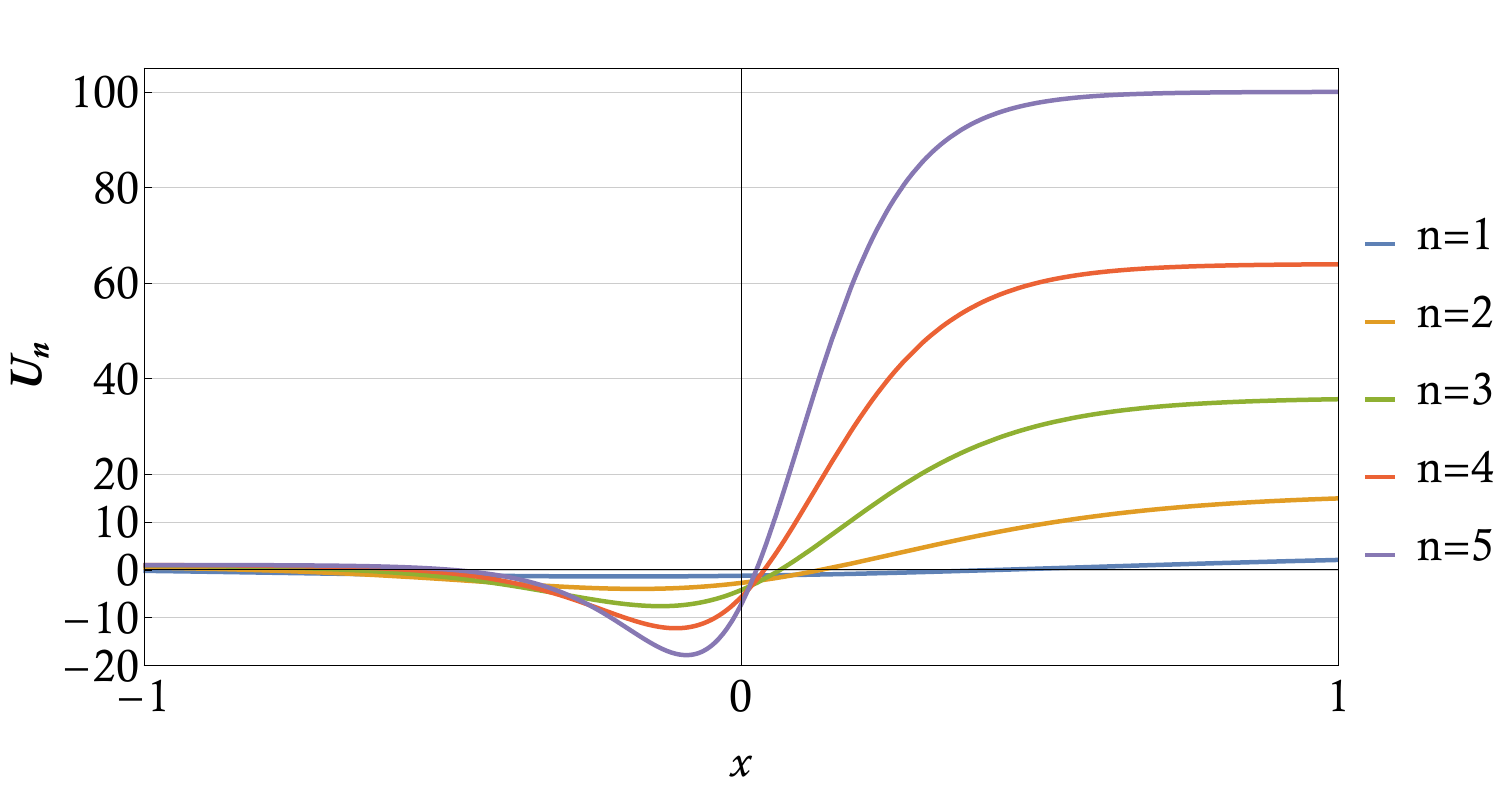}
\qquad
\includegraphics[width=.8\textwidth]{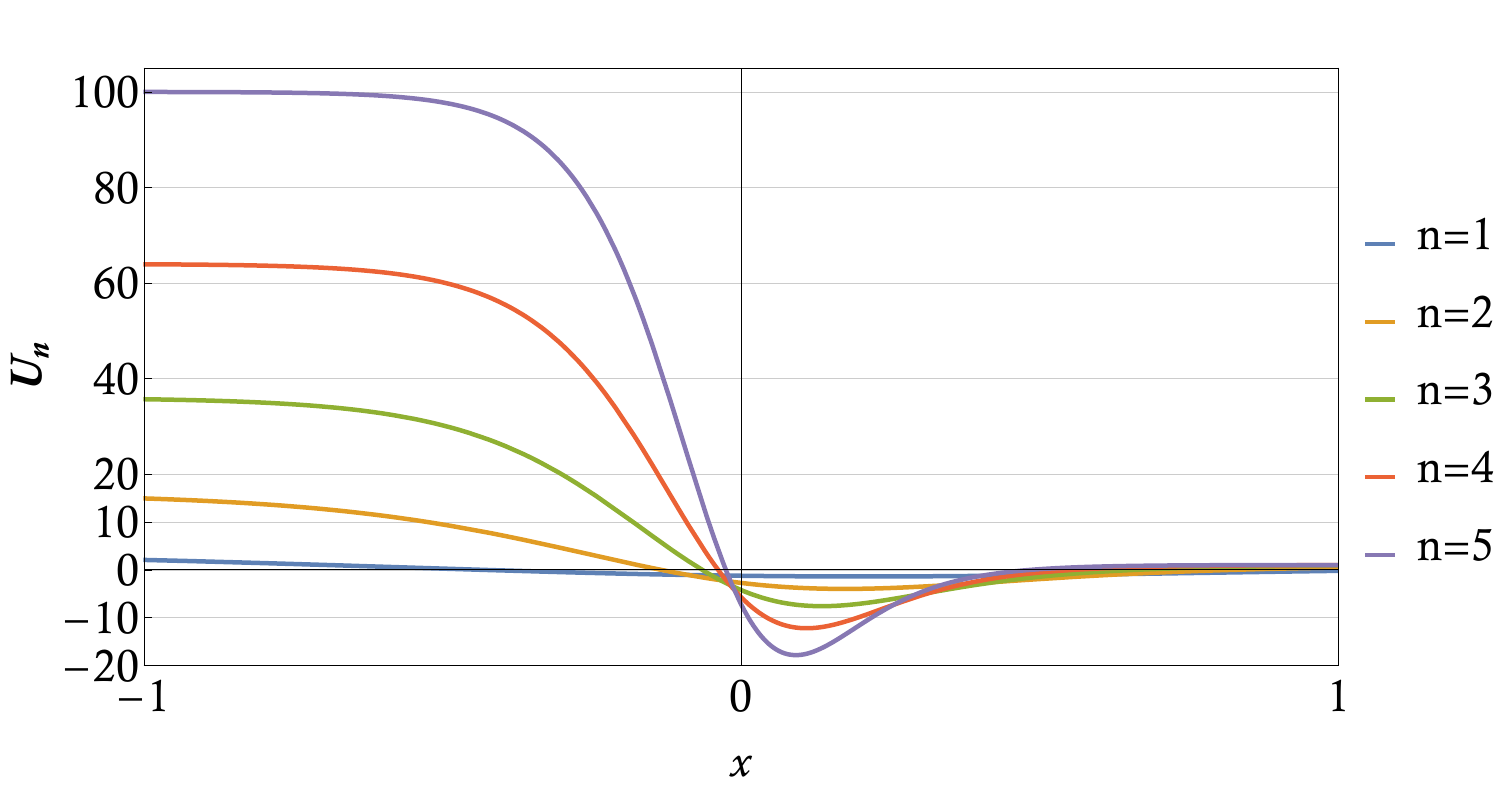}
\qquad
\caption{Examples of the Schr\"{o}dinger-like potential for the single static kink (top) and antikink (bottom).
\label{UK}}
\end{figure}

We consider a scalar field theory in 1+1 spacetime with a Lagrangian:
\begin{equation}
    \begin{aligned}
        \mathcal {L}=\int dx\left(\frac{1}{2}\partial_\mu\partial^\mu \phi-\frac{k^2}{2}\phi^2(\phi^{2n}-1)^2\right).
    \end{aligned}
\end{equation}
Notice that at $n=1$, the theory returns to the $\phi^6$ Theory. As shown in Fig. \ref{U}, the potential approaches a quadratic potential at $n\rightarrow \infty$. This allows us to examine the vacuum oscillation at vacuum $\phi=0$ and obtain an analytical solution at $n\rightarrow \infty$ limit. The equation of motion is:
\begin{equation}
\label{eq:1}
\begin{aligned}
\ddot {{\phi}}-{\phi}''+k^2\phi(\phi^{2n}-1)^2+k^2(\phi^{2n}-1)2n\phi^{2n+1}=0.\\
\end{aligned}
\end{equation}
The kink(antikink) solution can be explicitly integrated as
\begin{equation}
\label{eq:2}
    \begin{aligned}
    \phi_{n}=\frac{1}{(1+e^{\pm2nk(x-x_0)})^{\frac{1}{2n}}}.
    \end{aligned}
\end{equation}
At the $n\rightarrow \infty$ limit, the kink(antikink) solutions become a half compacton:
\begin{equation}
    \begin{aligned}
      \underset {n\rightarrow\infty}{Limit}\ \phi_{nK}=&1,x>0\\
&e^{kx},x<0.
    \end{aligned}
\end{equation}
 Figs. \ref{KF} shows some configurations of kink and antikink with their compact side at the origin
The fluctuations around the kink solution $\phi_s$ can be obtained from the linear perturbation. We set:
\begin{equation}
\label{eqnp}
    \begin{aligned}
       \widetilde\phi=\phi_{nK}+\eta e^{i \omega t},
    \end{aligned}
\end{equation}
The corresponding perturbative equation is 
\begin{equation}
    \begin{aligned}
     -   \eta''+U_{n}(x)\eta=0,
    \end{aligned}
\end{equation}
where $U_{n}(x)$ is the Schr\"{o}dinger-like potential. Substituting the expression of kink solution(  \ref{eq:2}) into $\phi_{s}$, the $U_{n}(x)$ can be written out explicitly as:

\begin{equation}
\label{eqnk}
    \begin{aligned}
 U_{n}(x)=1-\frac{(2 n+1) \left(2 (n+1) e^{-2 n (x-x_0)}-1\right)}{\left(e^{-2 n (x-x_0)}+1\right)^2}.
    \end{aligned}
\end{equation}
We plot the Schr\"{o}dinger-like potential of a kink and antikink in Fig. \ref{UK}. Notice that on one side, the lower mass threshold $m=1$ is independent of $n$, while on the other side, the higher mass threshold grows without an upper bound as $n$ increases.

\begin{figure}[htbp]
\centering
\includegraphics[width=.8\textwidth]{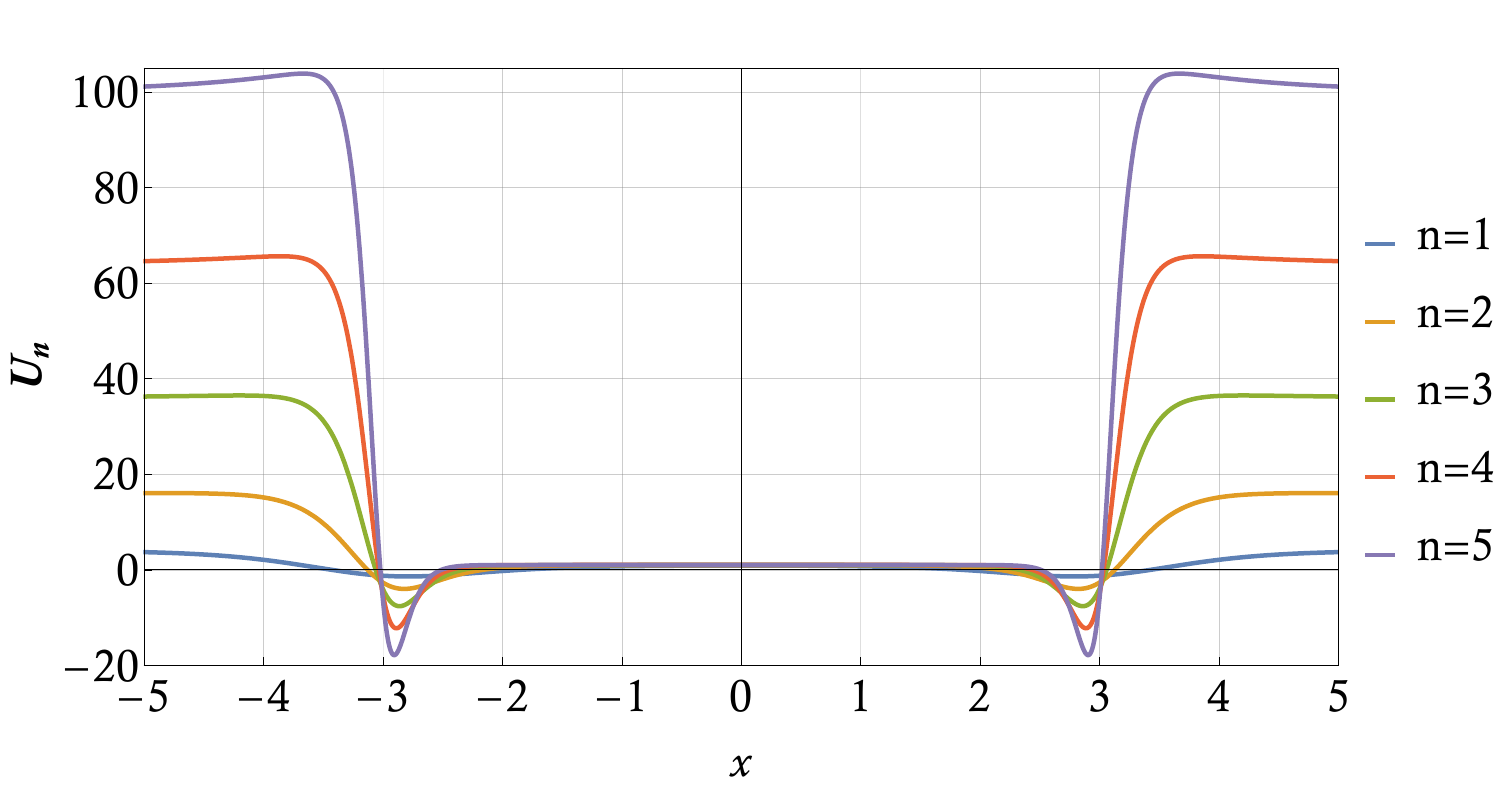}
\qquad
\includegraphics[width=.8\textwidth]{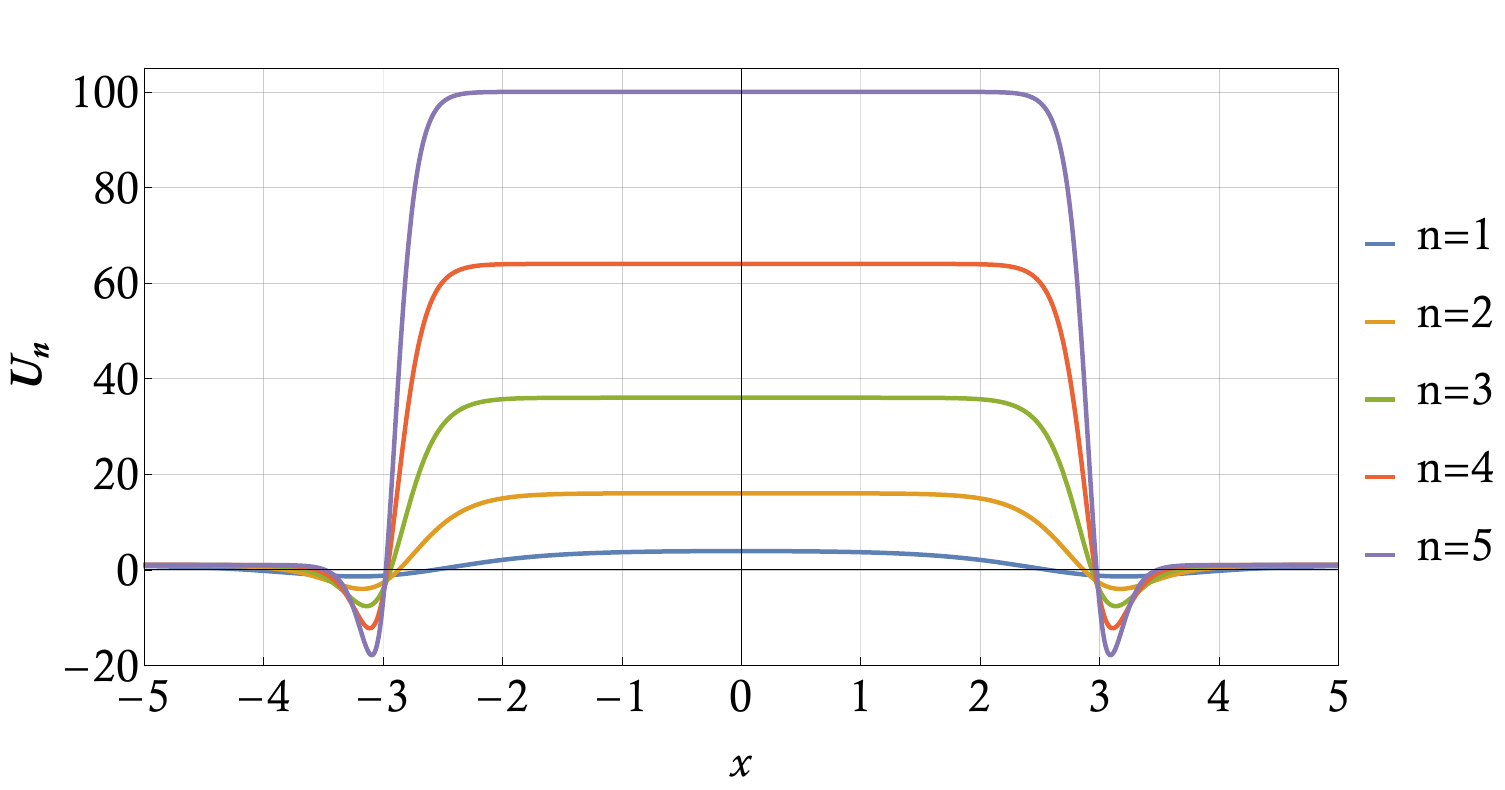}
\qquad
\caption{ The Schr\"{o}dinger-like potential for the superposition of the antikink-kink pair (top) and kink-antikink pair (bottom).
\label{Us}}
\end{figure}

Next, we consider $\phi_s$ as a superposition of an antikink and a kink. Fig. \ref{Us} shows the Schr\"{o}dinger-like potential for the antikink-kink pair and the kink-antikink pair with the half-separation $x_{0}=3$. In the antikink-kink case, the potential well tends to be a one-dimensional infinite square well plus two narrow wells near the boundaries, resulting in more vibration modes as $n$ increases. In the kink-antikink case, the height of the central plateau gets larger as $n$ increases.

\section{PDE simulation of kink antikink collision}
\label{PDE}
\subsection{KAK scattering}
In this section, we discuss kink collisions in the topological sector $\{0,1\}+\{1,0\}$. For the kink-antikink collision, the Pseudo-Spectral method is applied to discretize the space, and the Implicit Runge-Kutta method is applied to calculate the ODE in the time direction.  The initial half-separation $a_0$ is set to be $10$ and the spatial interval is $x\in[-220,220]$. To absorb the radiation at the boundary, we include an Error function damping term in the region $x\in[-220,-200]$ and $x\in[200,220]. $ 

The initial conditions for the kink-antikink scattering simulation are:
\begin{equation}
    \begin{aligned}
&\phi_n(x,0)=\phi_{nK}{(\gamma(x+a_{0}),0)}+\phi_{n\bar K}{(\gamma(x-a_{0}),0)}-1\\
&\dot{\phi_n}(x,0)=\dot\phi_{n K}{(\gamma(x+a_{0}),0)}+\dot\phi_{n\bar K}{(\gamma(x-a_{0}),0)},\\
    \end{aligned}
\end{equation}
where $\gamma\equiv 1/\sqrt{1-v^2}$ is the Lorentz factor. The solutions for the kink and antikink, denoted by $\phi_{nK}$ and $\phi_{n\bar{K}}$ respectively, are parameterized by $n$.

\begin{figure}[htbp]
\centering
\includegraphics[width=.8\textwidth]{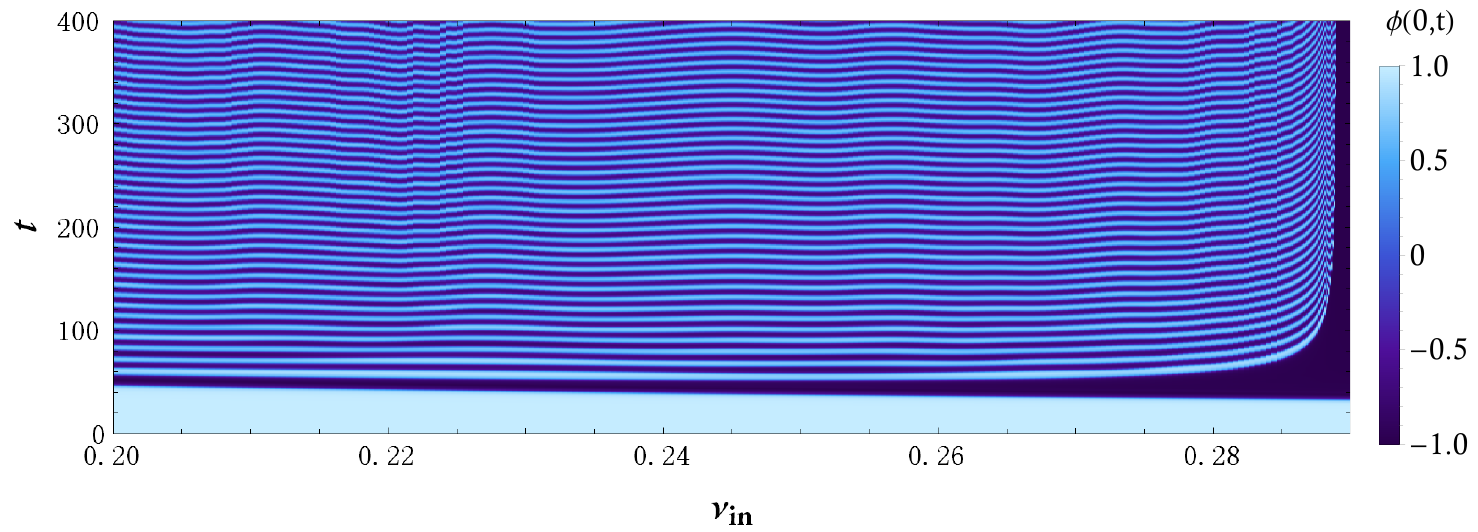}
\quad
\caption{KAK scattering: the fractal structure for $n=1$\label{fig:1}}
\end{figure}

\begin{figure}[htbp]
\centering
\includegraphics[width=.45\textwidth]{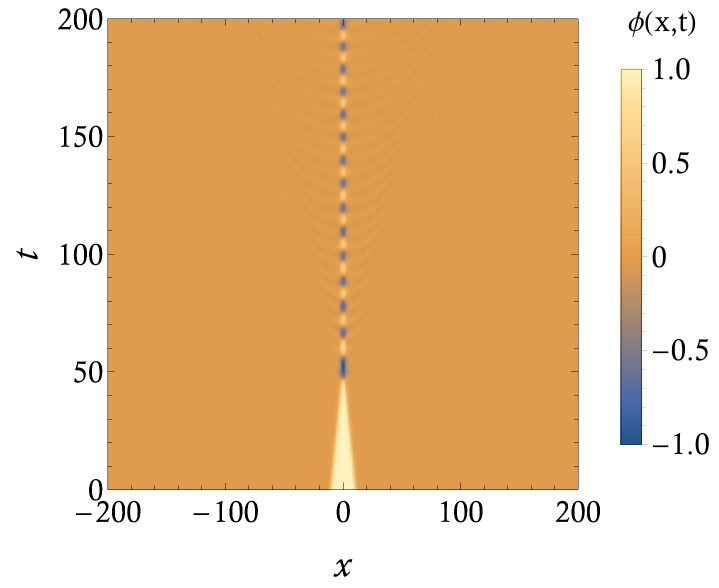}
\qquad
\includegraphics[width=.45\textwidth]{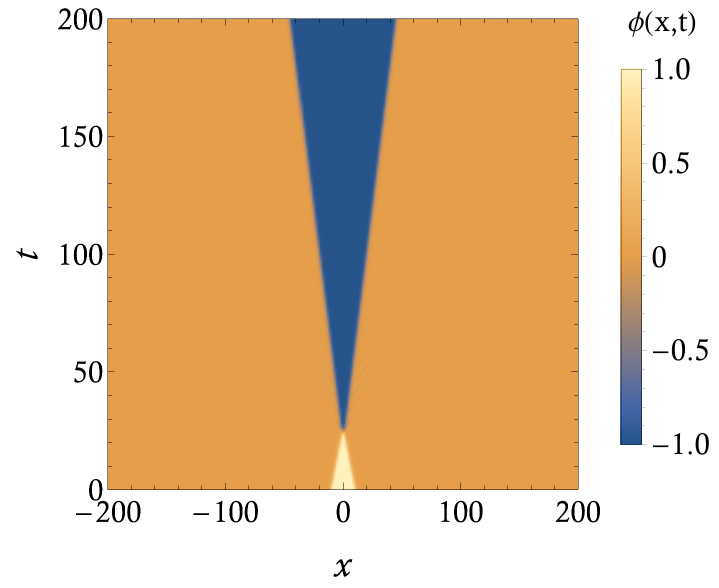}
\caption{Kink-antikink collision for $n=1$ with Left: $v=0.2$-bion state; Right: $v=0.4$-one bounce scattering\label{fig:2}}
\end{figure}

KAK scattering radiates stronger for larger $n$. For $n=1$, the half-compact model returns to the $\phi^6$ model\cite{dorey2011kink}. The fractal structure for KAK scattering is plotted in Fig. \ref{fig:1}. No resonance windows are observed. If the incident velocity is larger than the critical velocity $v_{c}\approx0.289$, kink and antikink could pass through each other. On the other hand, when the incident velocity is below the critical velocity, kink and antikink will annihilate and form a long-lived bion state(See Fig. \ref{fig:2}).

\begin{figure}[htbp]
\centering
\includegraphics[width=.8\textwidth]{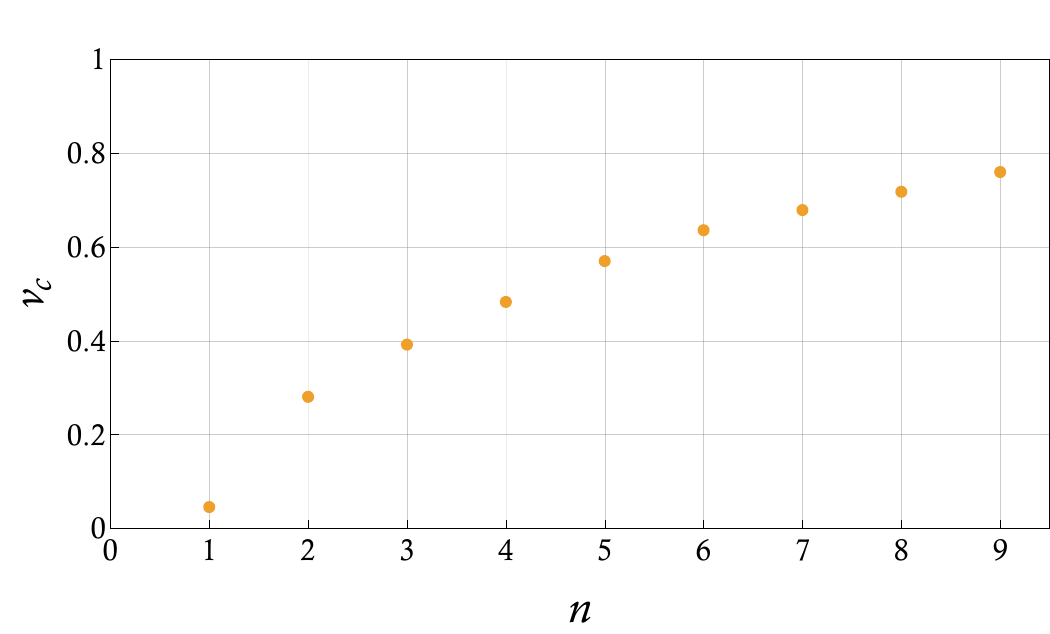}
\caption{The dependence of the critical velocity versus $n$. Some $v_{c}$ values are  0.705, 0.836, 0.894, 0.934, 0.962, 0.978, and 0.987 for $n$ values of 2, 3, 4, 5, 6, 7 and 8, respectively. \label{fig:3}}
\end{figure}

The value of $v_{c}$ increases rapidly for the first few $n$ and approaches the speed of light asymptotically for larger $n$.  The dependence on $v_{c}$ on $n$ is illustrated in Fig. \ref{fig:3}.

\begin{figure}[htbp]
\centering
\includegraphics[width=.8\textwidth]{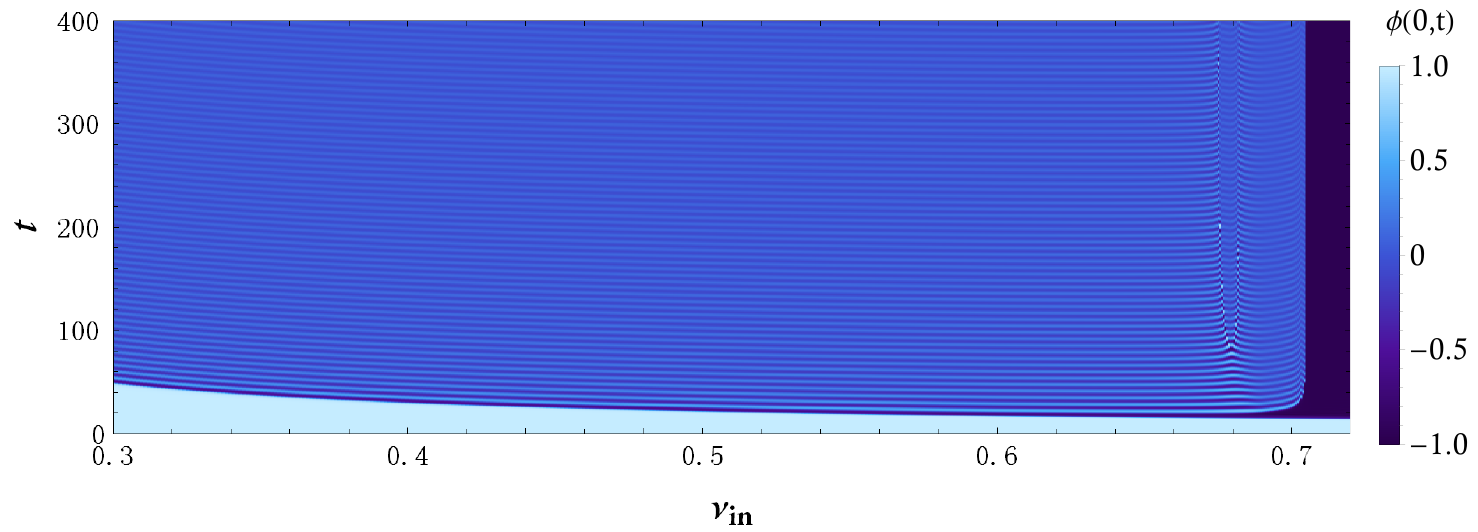}
\quad
\caption{KAK scattering: the fractal structure for $n=2$\label{fig:4}}
\end{figure}

In the case of $n=2$, the fractal structure is shown in Fig. \ref{fig:4}. A significant feature compared to the $\phi^6$ KAK scattering is the absence of bion.  As shown in Fig. \ref{fig:3}, in the interval $0<v_{c}<0.67$, kink and antikink annihilate. The peaks of the radiation move hyperbolically  and the boundary of the radiation moves at the speed of light. Similar behavior is also found in $\phi^8$ theory\cite{Exo}. 
\begin{figure}[htbp]
\centering
\includegraphics[width=.45\textwidth]{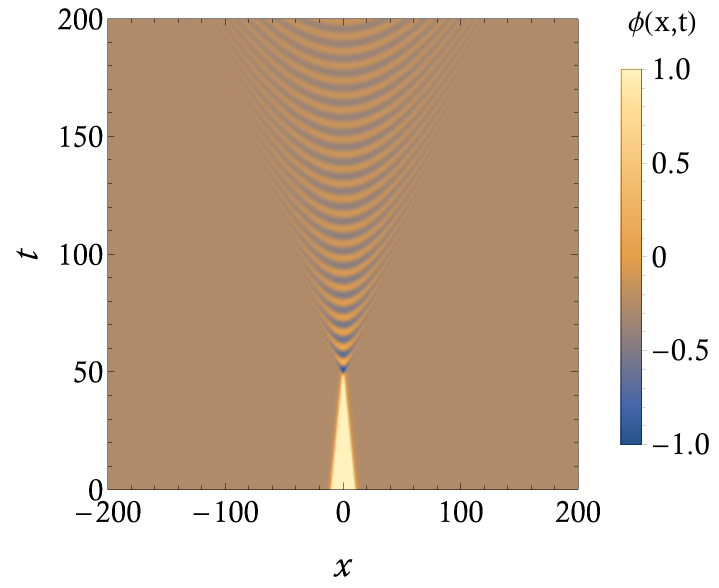}
\qquad
\includegraphics[width=.45\textwidth]{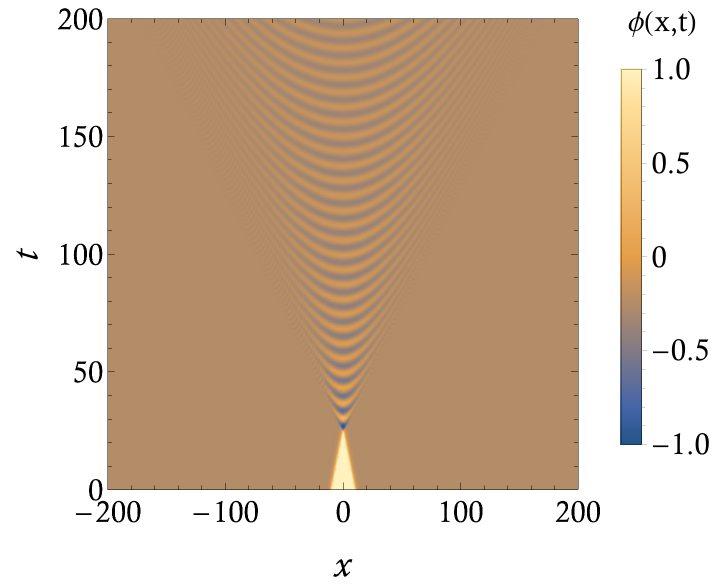}
\caption{Two dacay states formed in Kink-antikink collision for $n=2$ with Left: $v=0.2$; Right: $v=0.4$.\label{fig:5}}
\end{figure}

\begin{figure}[htbp]
\centering
\includegraphics[width=.8\textwidth]{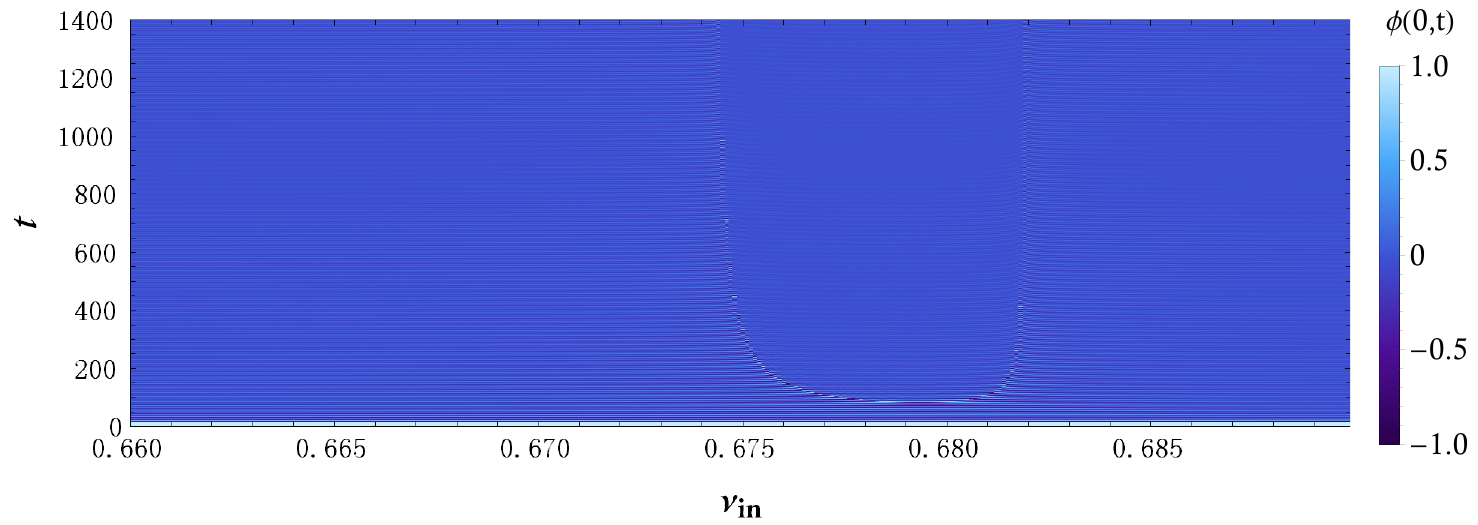}
\qquad
\includegraphics[width=.7\textwidth]{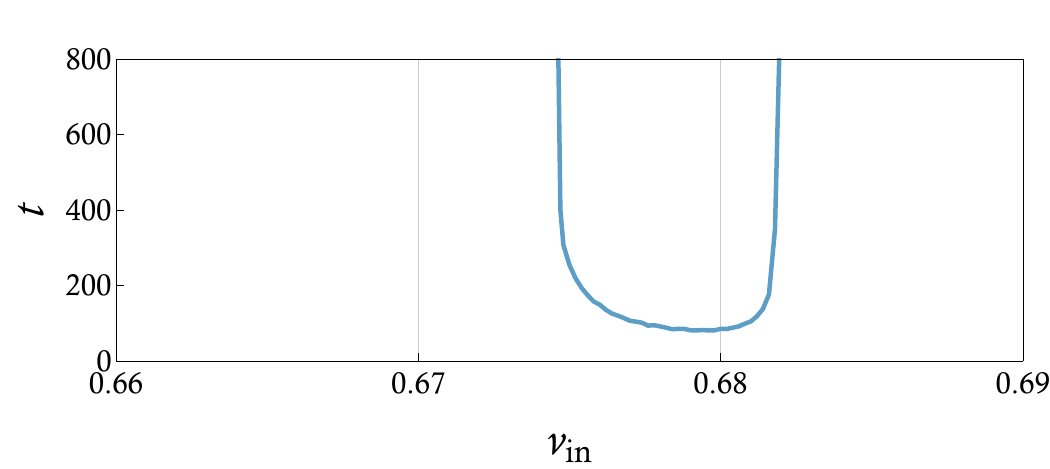}
\caption{Top: Zoomed in region for $0.660<v_{in}<0.690$ in KAK scattering for $n=2$. Bottom: the lifetime of the oscillons.\label{La}}
\end{figure}

\begin{figure}[htbp]
\centering
\includegraphics[width=.45\textwidth]{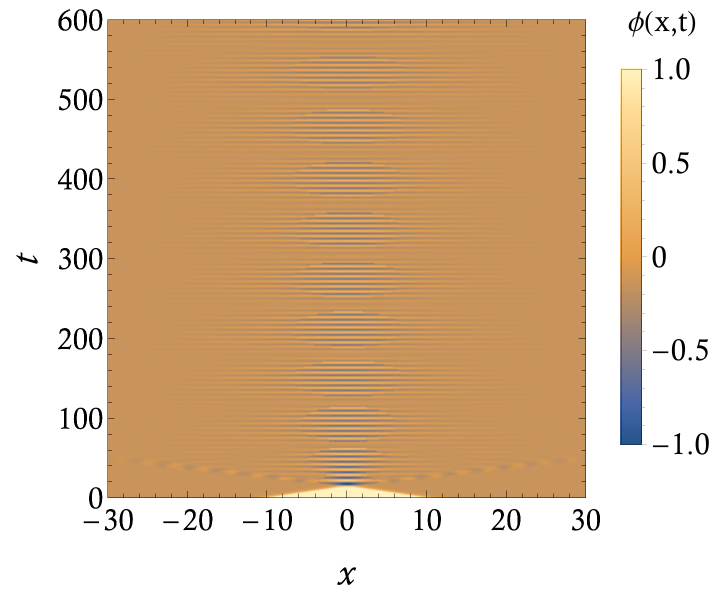}
\qquad
\includegraphics[width=.45\textwidth]{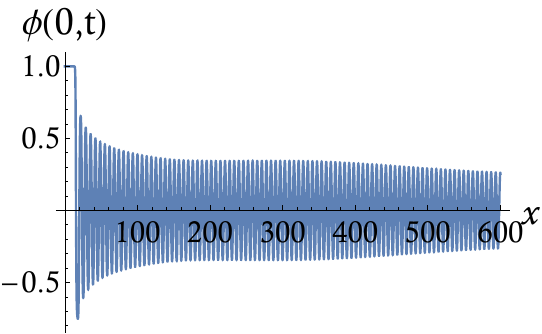}
\qquad
\includegraphics[width=.45\textwidth]{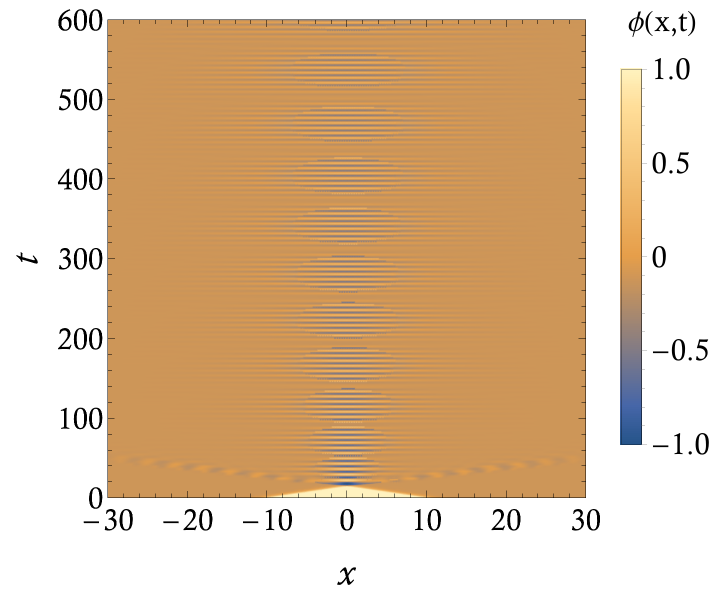}
\qquad
\includegraphics[width=.45\textwidth]{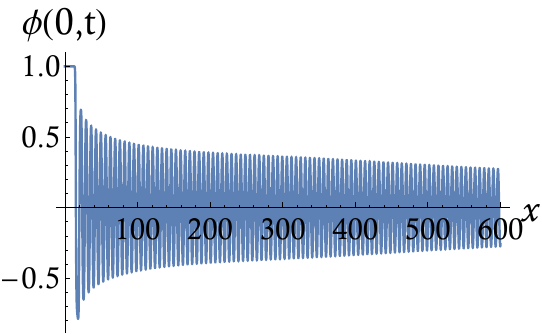}
\caption{KAK scattering for $n=2$: The field evolution for $v_{in}=0.6820$ (top left) and $v_{in}=0.6745$ (bottom left). The value of the field at the origin for $v_{in}=0.6820$ (top right) and $v_{in}=0.6745$ (bottom right).
\label{CO}}
\end{figure}

However, in the interval $0.6745<v_{in}<0.6820$, the bion doesn't decay completely. In Fig. \ref{La}, We find that the field configuration could form a short-lived oscillon. For the critical excited velocities $v=0.682$ and $v=0.6745$, the lifetimes of the formed oscillons are extremely long, suggesting their lifetimes are infinite. The two excited oscillons have different properties from $\phi^6$ bions. The width of $\phi^6$  bion remains approximately constant. However, for $n=2$, the widths of the two long-lived oscillons broaden as time increases, and their energy continues to disperse after the first collision. Fig. \ref{CO} shows the field evolution at the origin of the two oscillons for $n=2$.

\begin{figure}[htbp]
\centering
\includegraphics[width=.45\textwidth]{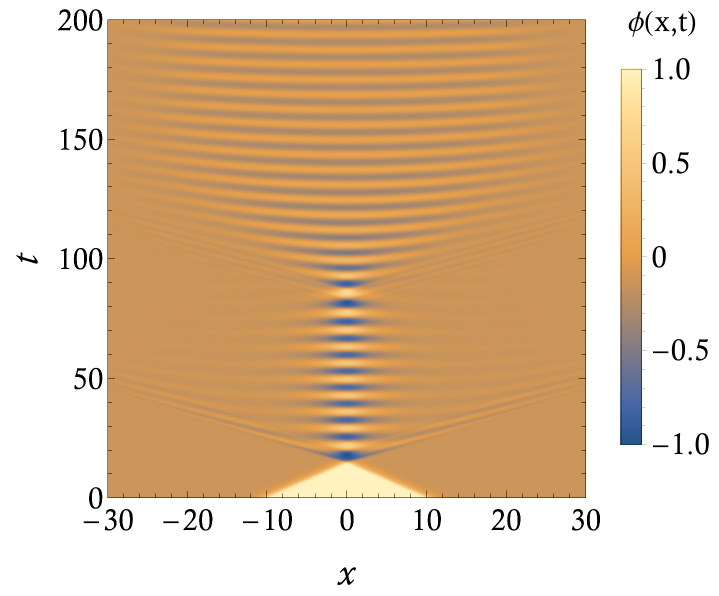}
\qquad
\includegraphics[width=.45\textwidth]{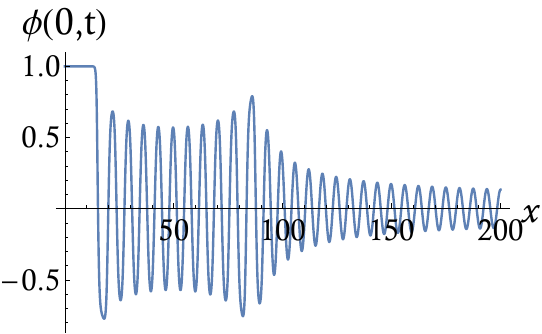}
\qquad
\caption{An excited oscillon with finite lifetime for $n=2$ and $v_{in}=0.68$. Left: the evolution of the field. Right: the value of the field at the origin of the oscillon.
\label{OSL}}
\end{figure}

Within the oscillating window $0.6745<v_ {in}<0.6820$, excited oscillons exhibit finite lifetimes.  Fig. \ref{OSL} illustrates the evolution of the oscillon and its value at $t=0$ with $v_{in}=0.68$. The oscillon forms after the first impact and keeps decreasing in amplitude; until about $t=55.6$, the amplitude starts to increase, reaching a maximum at $t=92.1$. Afterward, the oscillon collapses into a decayed state. All these collapse times are composed into the white "U" shape line in the oscillating window in Fig. \ref{La}.

\begin{figure}[htbp]
\centering
\includegraphics[width=.8\textwidth]{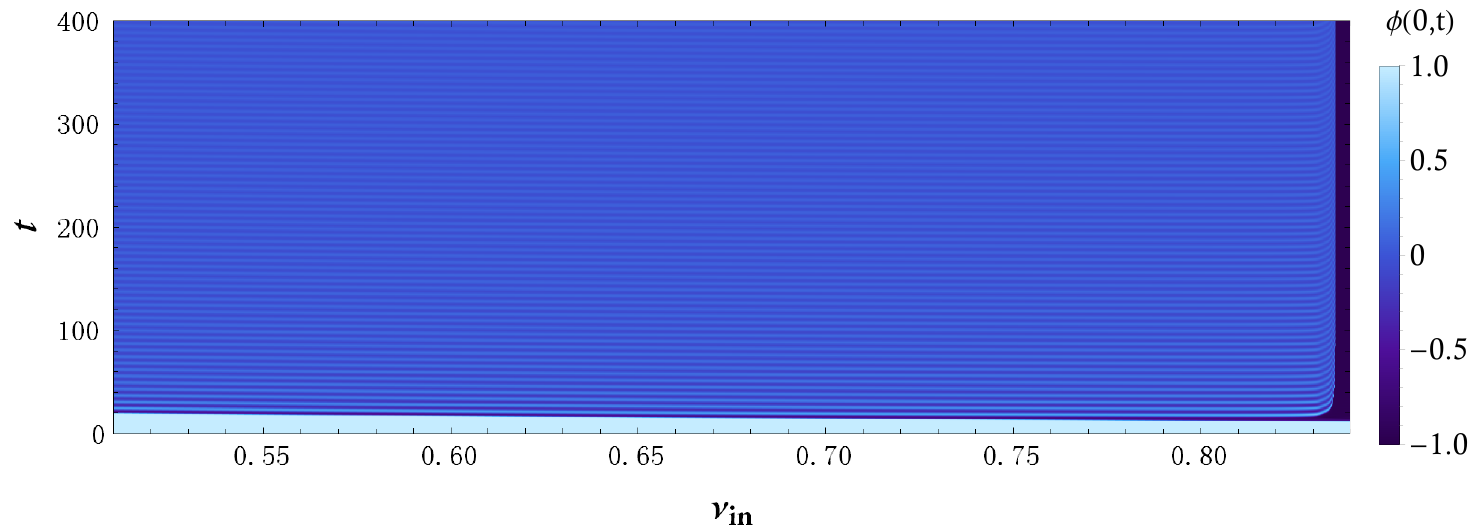}
\qquad
\caption{KAK scattering: the fractal structure for $n=3$.
\label{fr34}}
\end{figure}

\begin{figure}[htbp]
\centering
\includegraphics[width=.45\textwidth]{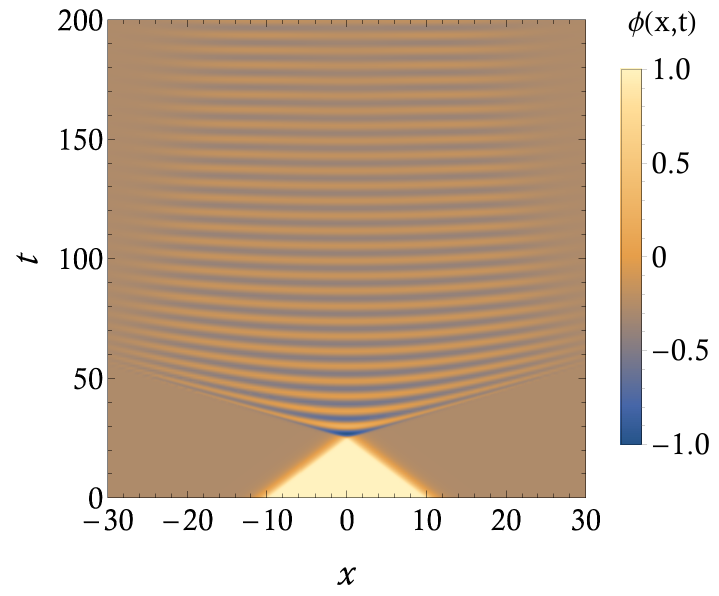}
\qquad
\includegraphics[width=.45\textwidth]{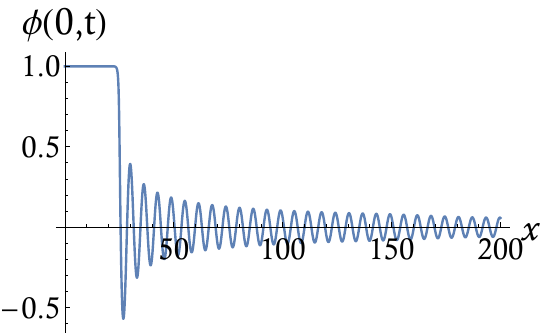}
\qquad
\caption{One example of decay state in KAK scattering for $n=3$ and $v_{in}=0.4$. Left: the evolution of the field.  Right: the field center value.
\label{OSI}}
\end{figure}

The emergence of oscillons with finite lifetimes can be regarded as a transition from non-compacton to compacton for the half-compact model with $n=2$. Fig. \ref{fr34} shows the fractal structure for $n=3$. No oscillon window is observed, since the compact property becomes more significant as n increases, making it harder for oscillons to form. As a result, all KAK scattering with incident velocities under the $v_c$ tends to decay completely. For instance, we plot the field configuration for  $v_{in}=0.4$ in Fig. \ref{OSI}.

\subsection{AKK scattering}
Now we discuss the antikink-kink collision in the sector $\{1,0\}+\{0,1\}$. We use the Pseudospectral method to discretize the space. In the time direction, we use the Implicit Runge-Kutta method for $n=1$ and $n=2$ and the LSODA method for $n>2$. The initial half-separation is set to be $a_{0}=10$. The initial conditions are
\begin{equation}
    \begin{aligned}
&\phi(x,0)=\phi_{nK}{(\gamma(x-a_{0}),0)}+\phi_{n\bar K}{(\gamma(x+a_{0}),0)}-1\\
&\dot{\phi}(x,0)=\dot{\phi}_{nK}{(\gamma(x-a_{0}),0)}+\dot\phi_{n\bar K}{(\gamma(x+a_{0}),0)}.\\
    \end{aligned}
\end{equation}

\begin{figure}[htbp]
\centering
\includegraphics[width=.8\textwidth]{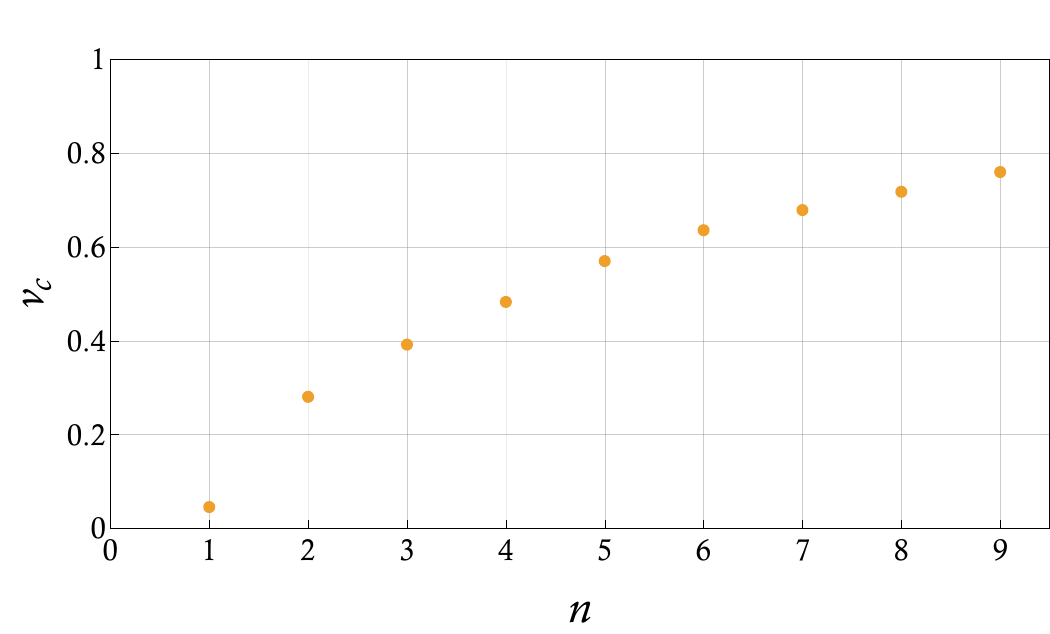}
\quad
\caption{The dependence of the critical velocity versus $n$. The critical velocities  $v_{c}$ are 0.046, 0.280, 0.392, 0.486, 0.580, 0.652, 
0.929, 0.964, and 0.981 for $n$ from 2 to 9, respectively.\label{AKKvc}}
\end{figure}

We first analyze the dependence of the critical velocity on $n$. In the AKK scattering, $v_{c}$ increases with $n$ and finally approaches the speed of light, which is the same as the KAK case. We test nine $n$'s and plot the dependence in Fig. \ref{AKKvc}. 

\begin{figure}[htbp]
\centering
\includegraphics[width=.8\textwidth]{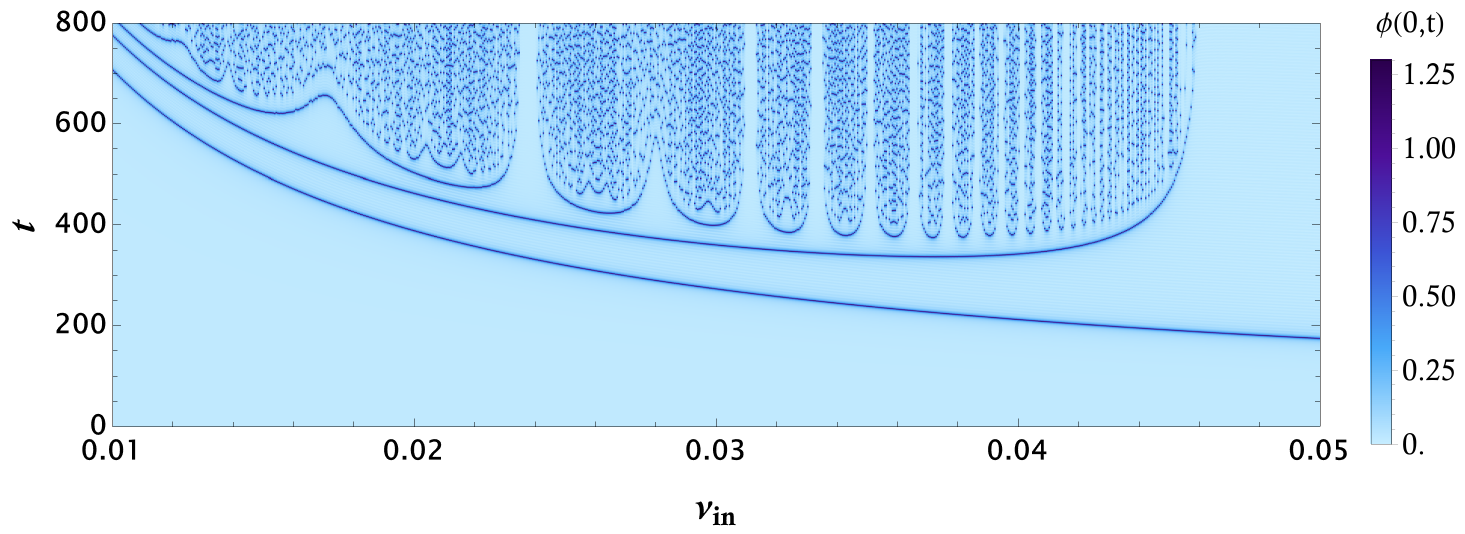}
\quad
\caption{AKK scattering: the fractal structure for $n=1$\label{1AKK}}
\end{figure}

For $n=1$, the half-compact model returns to the $\phi^6$ model. When $v_{in}>v_{c}$, the antikink and kink will pass through each other and move toward infinity. When $v_{in}<v_{c}$, the kink and antikink will form a bion state. However, for certain velocity intervals, they will separate after $k$ bounces, and the intervals are called k-bounce windows. In $\phi^6$ theory, the higher bounce windows are nested in the lower bounce windows. This pattern continues for all $k$, which composes the fractal structure. The resonance bounces come from the energy transfer between the vibrational modes and the translational mode \cite{campbell}. However, in the $\phi^6$ theory, the resonance fractal structure (see Fig. \ref{1AKK}) is triggered by the delocalized modes associated with the superposition of the antikink and kink.  \cite{dorey2011kink}.

\begin{figure}[htbp]
\centering
\includegraphics[width=.8\textwidth]{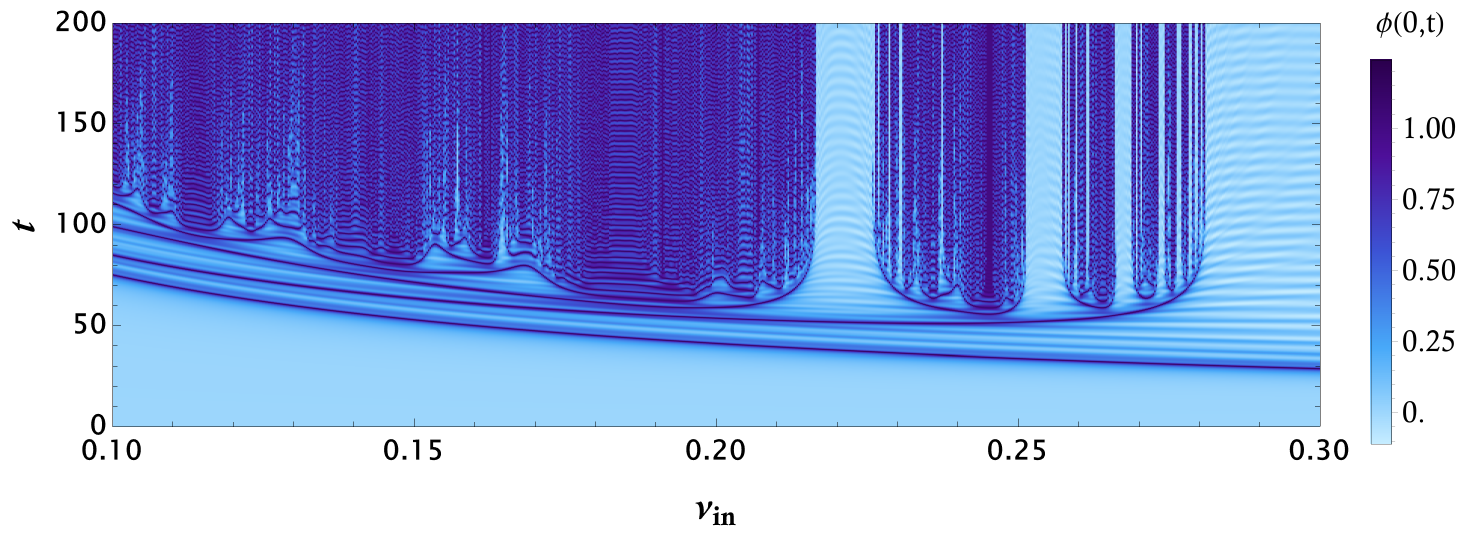}
\quad
\caption{AKK scattering: the fractal structure for $n=2$\label{2AKK}}
\end{figure}

\begin{figure}[htbp]
\centering
\includegraphics[width=.28\textwidth]{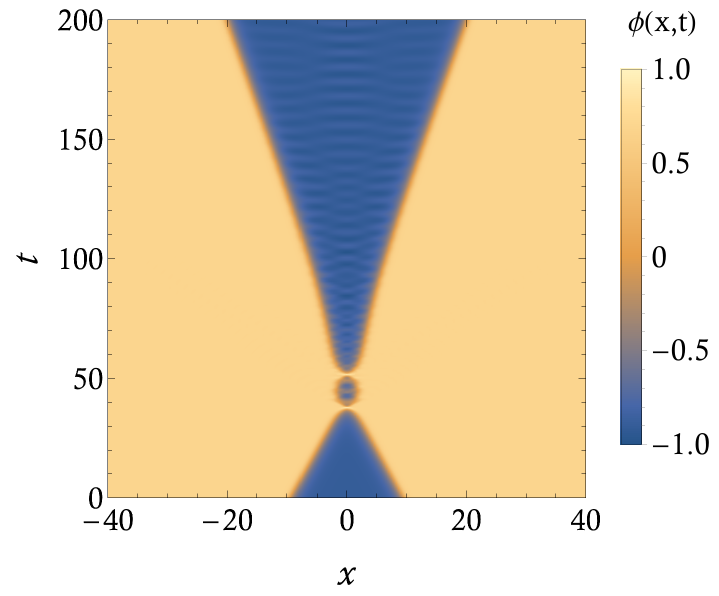}
\qquad
\includegraphics[width=.28\textwidth]{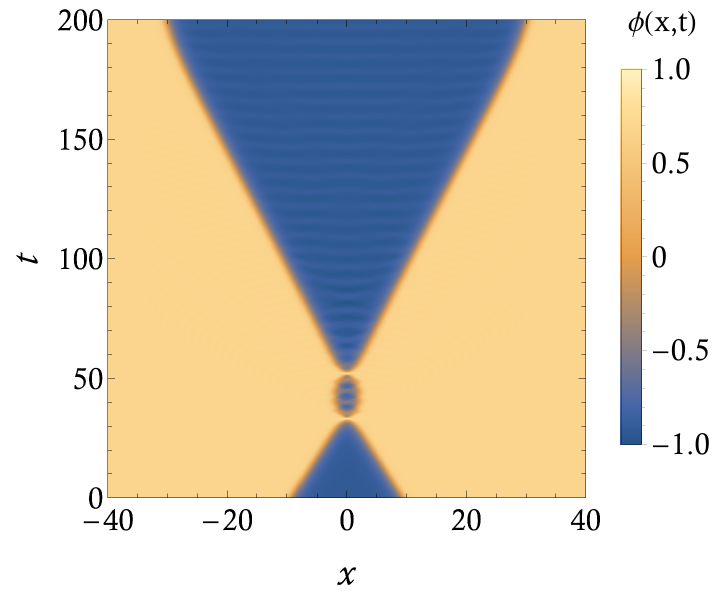}
\qquad
\includegraphics[width=.28\textwidth]{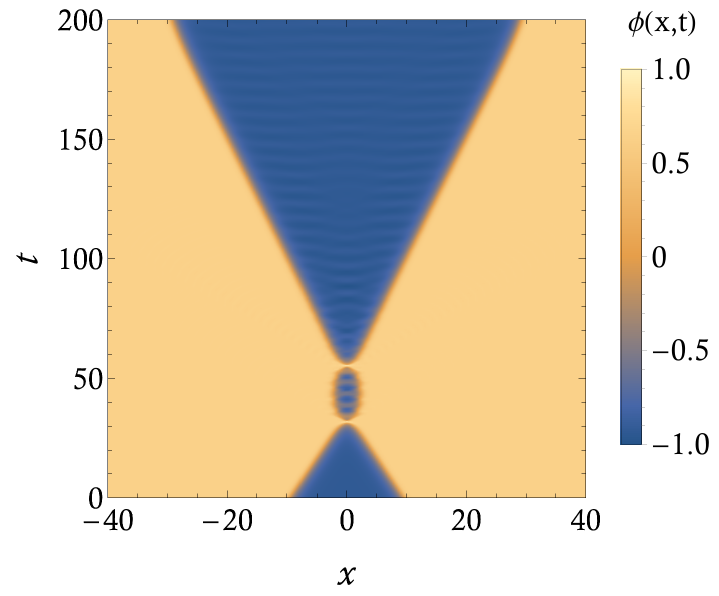}
\qquad
\caption{The evolution of the field of the first three 2-bounce windows in AKK scattering for $n=2$. Left: $v_{in}=0.22$,  Middle: $v_{in}=0.255$,  Right: $v_{in}=0.267$.
\label{22BW}}
\end{figure}

\begin{figure}[htbp]
\centering
\includegraphics[width=.28\textwidth]{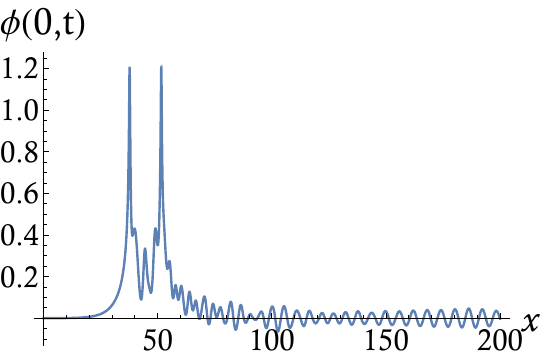}
\qquad
\includegraphics[width=.28\textwidth]{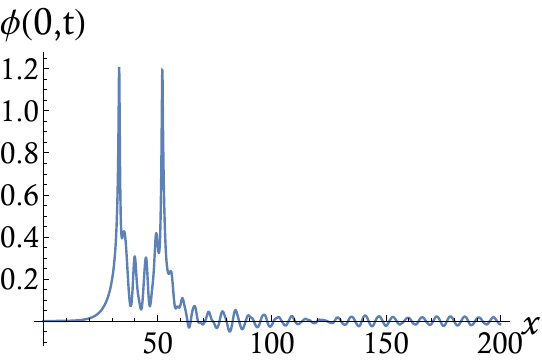}
\qquad
\includegraphics[width=.28\textwidth]{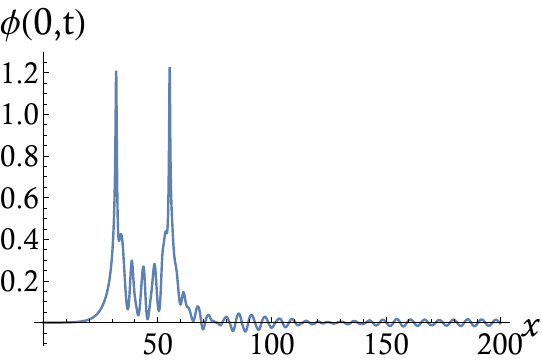}
\qquad
\caption{The field center values of the first three 2-bounce windows in AKK scattering for $n=2$. Left: $c_{in}=0.22$, $n=4$,  Middle: $v_{in}=0.255$, $n=5$,  Right: $v_{in}=0.267$, $n=6$.
\label{2-2BW}}
\end{figure}

For $n=2$, we plot the fractal structure in Fig. \ref{2AKK} . Fig. \ref{22BW} shows the evolution of the field for the first three 2-bounce windows. The order of 2-bounce windows can be detected from the number of small oscillations $N$ between the first and the second impact. In Fig. \ref{2-2BW}, we plot the value of the field at the origin for these bounce windows with $n= 4, 5, 6$, respectively. 

\begin{figure}[htbp]
\centering
\includegraphics[width=.8\textwidth]{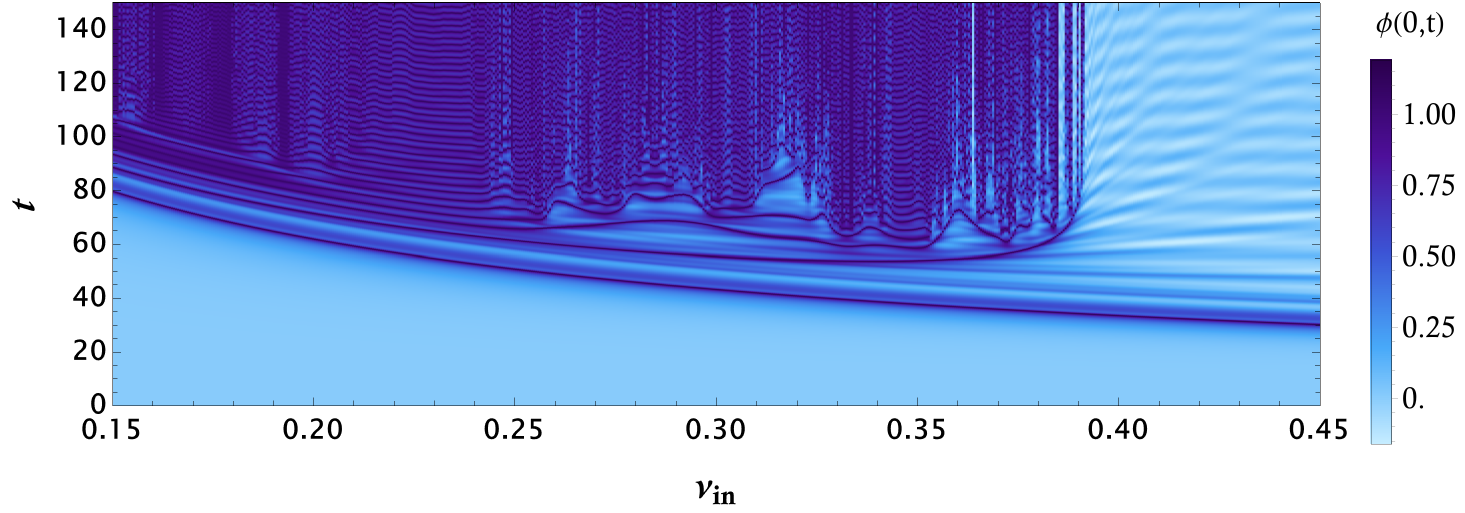}
\quad
\caption{AKK scattering: the fractal structure for $n=3$\label{3AKK}}
\end{figure}

\begin{figure}[htbp]
\centering
\includegraphics[width=.28\textwidth]{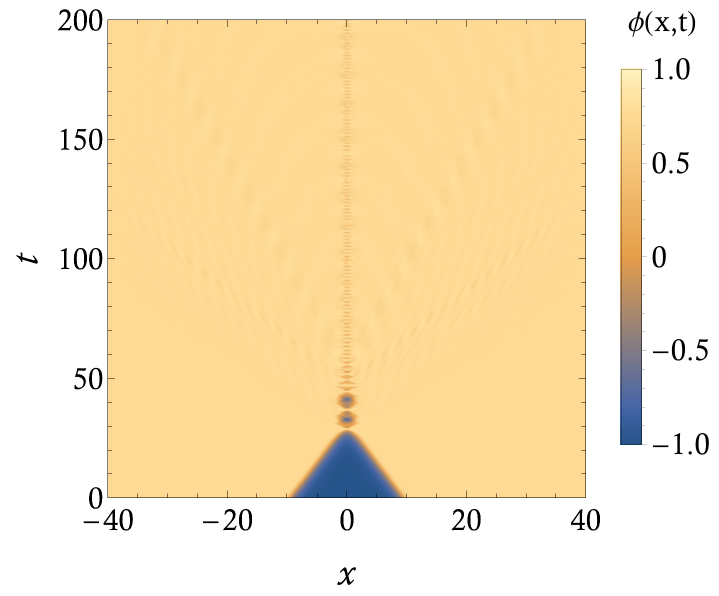}
\qquad
\includegraphics[width=.28\textwidth]{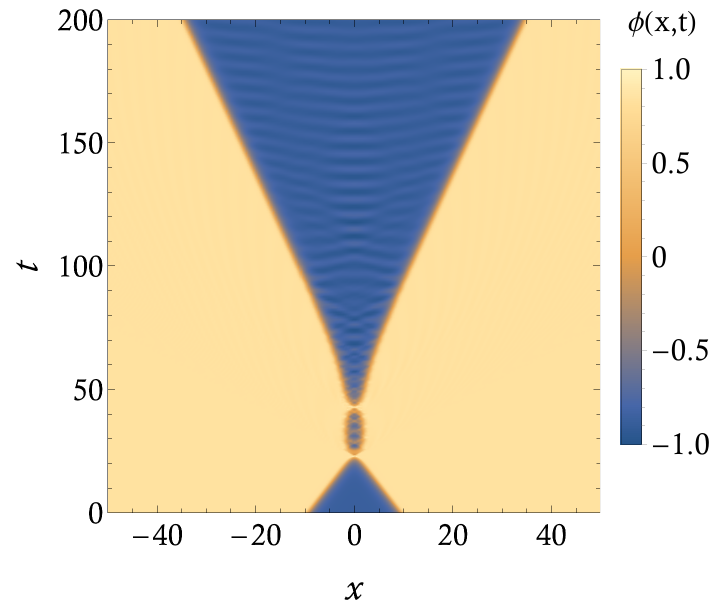}
\qquad
\includegraphics[width=.28\textwidth]{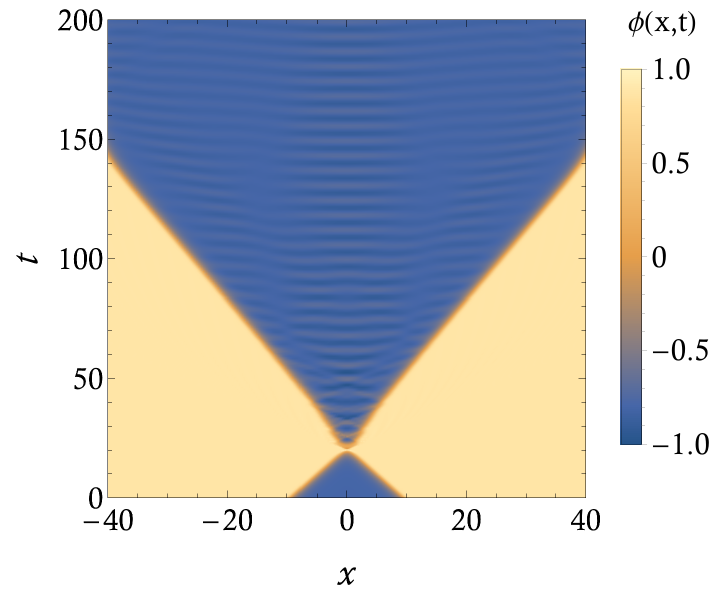}
\qquad
\caption{The evolution of the field in AKK scattering for $n=3$. Left: $v_{in}=0.300$-bion state,  Middle: $v_{in}=0.387$-2-bounce window,  Right: $v_{in}=0.452$-inelastic scattering.
\label{3W}}
\end{figure}

\begin{figure}[htbp]
\centering
\includegraphics[width=.28\textwidth]{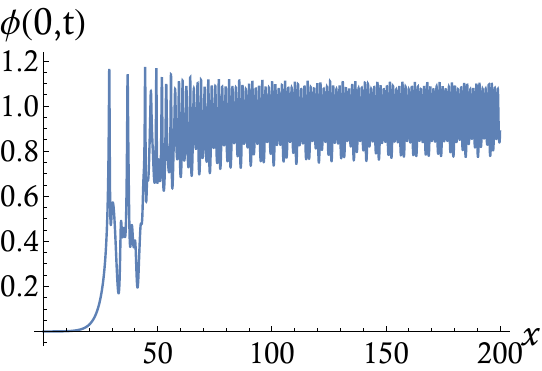}
\qquad
\includegraphics[width=.28\textwidth]{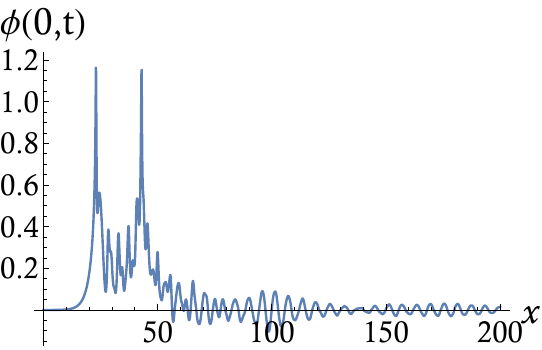}
\qquad
\includegraphics[width=.28\textwidth]{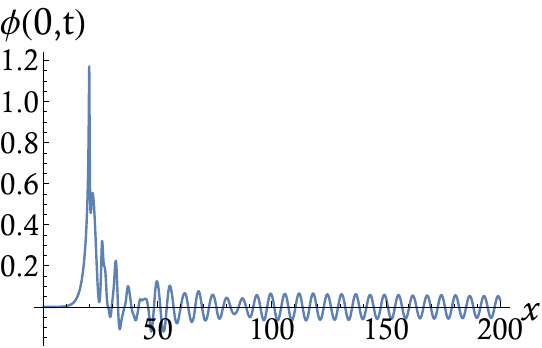}
\qquad
\caption{The center field value in AKK scattering for $n=3$. Left: $v_{in}=0.300$-bion state,  Middle: $v_{in}=0.387$-2-bounce window,  Right: $v_{in}=0.452$-inelastic scattering.
\label{3-W}}
\end{figure}

The fractal structure for $n=3$ is shown in Fig. \ref{3AKK}. Some examples of scattering are shown in Fig. \ref{3W} and Fig. \ref{3-W}. In the left panel, the antikink and kink annihilate and a bion state is formed after the first impact. In the middle panel, a 2-bounce window occurs. In the right panel, with enough kinetic energy, the antikink and kink get reflected after their first bounce. 

Compared to the $n=2$ case, the 2-bounce windows for the $n=3$ case become more narrow. We propose that the phenomenon may be related to the increasing delocalized modes. If the kink and antikink are not close enough, the Schr\"{o}dinger well(\ref{eqnk}) gets deeper as $n$ increases, resulting in an increase in the delocalized modes. In Fig. \ref{3-W}, the middle panel shows that the small oscillations are irregular, which implies that more than one vibration modes have been excited between the first and the second impact. The increase in vibration modes can lead to the vanish of 2-bounce windows, which is suggested by \cite{Adamrev6} and verified in a toy model proposed by \cite{LONG}. For larger $n$, resonance bounce phenomena will be suppressed more strongly.

\begin{figure}[htbp]
\centering
\includegraphics[width=.8\textwidth]{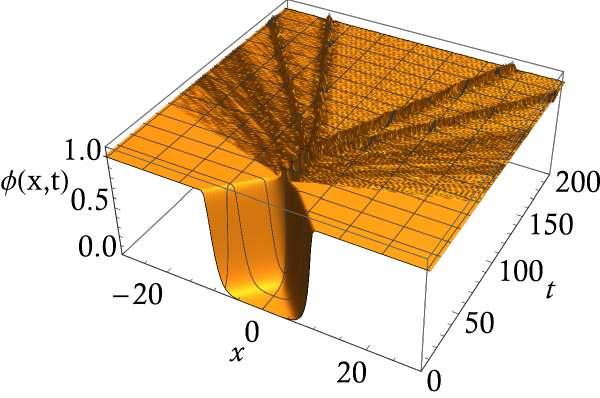}
\quad
\caption{The propagating oscillon for $n=3$ with $v_{in}=0.288$.
\label{po}}
\end{figure}

Propagating oscillons are observed in other higher polynomial models, such as some other compacton model \cite{compactoscillons}, \cite{Bazeia:2019tgt}, \cite{Bazeia:2020car} and the $\phi^8$ model \cite{shortrange}. 
This phenomenon is also observed in our model. Fig.\ref{po} shows an example for $n=3$ with an incident velocity $v=0.288$ located near the first 2-bounce window. After the first impact, antikink and kink form a temporary bion state and radiate within its lifetime. At the end of its lifetime near $t=60$, the bion state splits into two pairs of propagating oscillons. The pair of oscillons with a higher outgoing velocity oscillates weakly, while the pair with a lower velocity oscillates strongly. This phenomenon may be related to the extra internal or quasi-internal states of the kink pair.

\section{Analytic properties of radiation like solution}
\label{Analytic}
\subsection{Analytic solution at v=c}

As we have seen in Section \ref{themodel}, the potential approaches a quadratic form in the interval $\phi\in (-1,1)$. Therefore, in this section, we will exploit this partially analytical property of the potential to give a radiation-like one-parameter family of rogue wave solutions. The potential at $n\rightarrow \infty$ approaches a piecewise function:
    \begin{equation}
        \begin{aligned}
            \underset{n\rightarrow \infty}{Lim}V[\phi]=\begin{cases}
                \infty, &|\phi|>1\\
                    0,&|\phi|=1\\
                \frac{k^2}{2}\phi^2,&|\phi|<1.\\

            \end{cases}
        \end{aligned}
    \end{equation}

   If the field value is restricted to $|\phi|<1$, the equation of motion can be solved by the separation of variables. However, these solutions don't yield an interesting finite energy configuration. To obtain such solutions, we must transform the discontinuity of the field at $|\phi|=1$ into an extra boundary condition. To get this boundary condition, let's first rewrite $x,t$ into a single variable. Every value of this variable represents a curve in spacetime, and the boundary corresponds to some specific value of this variable.
 
   Let:
   \begin{equation}
       \begin{aligned}
\phi(x,t)=f(z), z=\frac{g(x,t)}{4} .
       \end{aligned}
   \end{equation}
The boundary condition at $|f(z)|=1$ will be matched after solving the general solution of the ODE in the bulk. Substituting this into the equation of motion, the derivative terms become:
    \begin{equation}
        \begin{aligned}
            \ddot{f}(z)-f''(z)=&\frac{1}{4}(\dot g(x,t)^2-g(x,t)'^2)\frac{d^2f}{dz^2}+\frac{1}{2}(\ddot g(x,t)-g''(x,t))\frac{df}{dz}\\
            &\dot g(x,t)^2-g'(x,t)^2=F(z)\\
            &\ddot g(x,t)-g''(x,t))=G(z),\\
        \end{aligned}
    \end{equation}
   
where $F(z)$ and $ G(z)$ are some functions of $z$. These nonlinear equations are hard to solve in general. Therefore, we have to use an Ansatz, which has been used to solve the Sine-Gordon equation \cite{PDEtoODE} and other Sine-Gordon-like models\cite{compactoscillons} \cite{shockwave}. The Ansatz is as follows:
        
    \begin{equation}
        \begin{aligned}
            z=g(x,t)=\frac{1}{4}(t^2-x^2), 
\phi(x,t)=f(g(z)).
        \end{aligned}
    \end{equation}

Substituting this into the PDE, the derivative terms become:
\begin{equation}
    \begin{aligned}   
\ddot{f}(z)-f''(z)&=\frac{1}{4}(t^2-x^2)\frac{d^2f}{dz^2}+\frac{1}{2}(1+1)\frac{df}{dz}\\
&=-k^2f(z)+k^2B(f(z)),
    \end{aligned}
\end{equation}
where $-k^2 f(z)$ is the bulk term and $B(f(z))$ is the term that provides a boundary at $n \rightarrow \infty$. As a summary, the equation of motion transforms to the following second-order ODE:
\begin{equation}
    \begin{aligned}
z\frac{d^2f}{dz^2}+\frac{df}{dz}=-k^2f(z)+k^2B(f(z)).\label{eom}
    \end{aligned}
\end{equation}
To solve this equation, we first ignore the boundary term $B(f(z))$. For $z>0$, let $\sqrt{z}\ k=\rho$, the equation\ref{eom} )  becomes: 
\begin{equation}
    \begin{aligned}
\frac{d^2f}{d\rho^2}+\frac{1}{\rho}\frac{df}{d\rho}+f(\rho)=0.
    \end{aligned}
\end{equation}

Which is the Bessel equation. The solution is:
\begin{equation}
    \begin{aligned}
        f(z)=c_1 J_0\left(2 k \sqrt{z}\right)+ c_2 Y_0\left(2 k \sqrt{z}\right),(z>0).
    \end{aligned}
\end{equation}

For $z<0$, let $i\sqrt{|z|
}k=\rho$,  the equation of motion then transforms into a modified Bessel equation
\begin{equation}
    \begin{aligned}
        \frac{d^2f}{d\rho^2}+\frac{1}{\rho}\frac{df}{d\rho}-f(\rho)=0.
    \end{aligned}
\end{equation} 
The solution is the Modified Bessel functions
\begin{equation}
    \begin{aligned}
       f(z)= c_1 I_0\left(2k \sqrt{-z}\right)+ c_2 K_0\left(2k \sqrt{-z}\right),(z<0).
    \end{aligned}
\end{equation}

With the solution in the bulk, it's time to consider the boundary term $B(f(z))$. To get some intuition for $B(f(z))$, we consider a Classical Mechanics analogy of equation (\ref{eom}).  This analogy is applied by treating z as t, f as x, and treating the boundary value problem as an initial value problem. The equation of motion for the field now becomes a Classical Mechanics system of an imaginary particle.
    \begin{equation}
        \begin{aligned}
t\frac{d^2x}{dt^2}+\frac{dx}{dt}=-k^2x(t)+k^2B(x(t)).\\
        \end{aligned}
    \end{equation}

The mass m=t, coming from the term $t\frac{d^2x}{dt^2}$, increases with time. At $t=0$, it forms a singularity. This singularity can be removed, since the Bessel function of the first kind and the Modified Bessel function of the first kind can both cross 0. For t<0, the mass becomes negative, which means all the force gets an extra minus sign. 

The potential is $\frac{k^2}{2}x^2$ for $|x|<1$ and will arrive at the discontinuities at $|x|=1$. The term -$\frac{df}{dz}$ shows this system has a damping. 
\begin{figure}
\centering
\includegraphics[width=.45\textwidth]{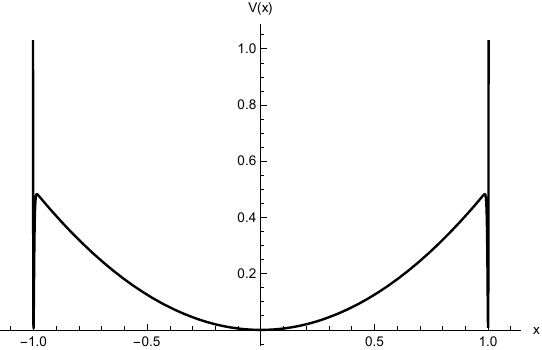}

\caption{potential V(x) at $n=200, k=1, t>0$}
\end{figure}

At $t<0$ the imaginary mass turns the damping term to a driving term, and the imaginary particle can only be stable at $x=\pm1$. We choose this trivial solution for $t<0$. For  $t>0$, the imaginary particle has two kinds of solutions. The first one is the trivial solution $x=1$. The second one is the oscillating x=0 solution. This oscillation produces a radiation-like solution in the original field system. 

\begin{figure}[htbp]
\centering
\includegraphics[width=.45\textwidth]{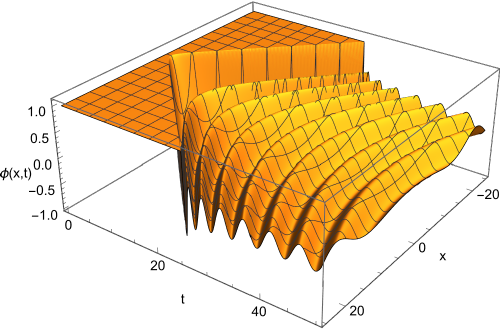}
\quad
\includegraphics[width=0.45\textwidth]{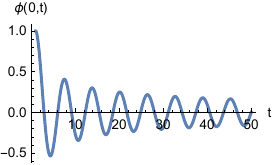}
\caption{
    Left: Solution at $z_0=1/4$ The solution  for $z>z_0$ is $\phi(x,t)=1.22713 J_0\left( \sqrt{t^2-x^2}\right)+0.69123 Y_0\left( \sqrt{t^2-x^2}\right)$\label{fig:8} \quad Right: $\phi(0,t)$ value of the right figure.\label{analytical}}\end{figure}

Now we go back to the field system. The value and the derivative of the two solutions at the boundary can be matched to be the same for $z>0$. The two solutions can be glued to one. Notice that the third-order derivative at the glued point is not the same, but the equation of motion only requires at most second-order derivative to be continuous. A similar example is Norton's dome\cite{Norton}, where the author shows the ill property when classical mechanics involves non-smooth potential. However, we don't need to worry about this non-smooth property in field theory.  Similar ambiguity also appears in the kink solution transforming from kink to a compacton. This ambiguity will be resolved for solutions with finite $n$. The only problem is that finding a solution at finite n in our case is much harder than in the kink case.  Therefore, in this work, we will only discuss the $n\rightarrow\infty$ case. The solution for our system is:
\begin{equation}
    \begin{aligned}
 f(z)=       
        \begin{cases}
            c_1 J_0\left(2 k \sqrt{z}\right)+ c_2 Y_0\left(2 k \sqrt{z}\right),z\geq z_0\\
            \\
            1,z<z_0,
        \end{cases}   \label{analytical sol}
    \end{aligned}
\end{equation}
where $f(z_0)=1$. $J_0$ and $Y_0$ are the Bessel functions of the first and the second kind, respectively. 
The $c_1$ and $c_2$ should satisfy extra conditions to keep the value and the derivative of the solution continuous:
\begin{equation}
     \begin{aligned}
         c_1 J_1(2z_0)+c_2 Y_1(2z_0)=0\\
c_1 J_0(2z_0)+c_2 Y_0(2z_0)=1.
     \end{aligned}
 \end{equation}
Fig. \ref{analytical} illustrates the configuration of the solution. The field remains constant for $t<2z_0$ and suddenly collapses into radiation with a boundary traveling faster than the speed of light. 
 The boundary curve of the solution is $4=t^2-x^2$, which means the exact solution has an ultra-relativistic property. This property possibly originates from the discontinuity of the potential. We will later show at a certain limit, these solutions can approach an antikink-kink scattering solution with a lightlike boundary.

\begin{figure}[htbp]\centering
\includegraphics[width=.6\textwidth]{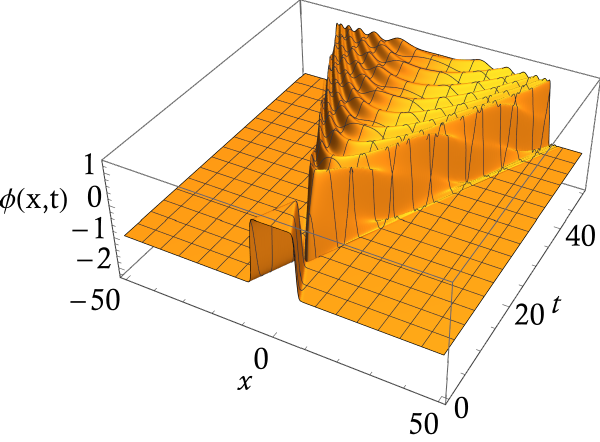}
\caption{
    $\phi^4$ AKK at $v=0.99$ \label{phi4}}\end{figure}

 In the bulk, the oscillation is ultra-relativistic too. However, similar behavior also appears in Section\ref{PDE} and in the kink collisions of other models\cite{Exo}\cite{shortrange}. Fig. \ref{phi4} shows even $\phi^4$ KAK collision has a similar structure. We suggest the group velocity might not be ultra-relativistic, and we will study the ultra-relativistic property in future work.

\begin{figure}
\centering
\includegraphics[width=.45\textwidth]{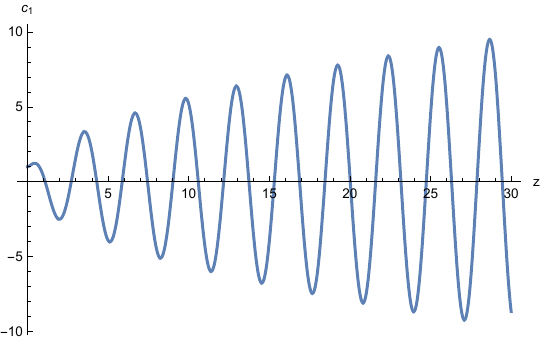}
\quad
\includegraphics[width=0.45\textwidth]{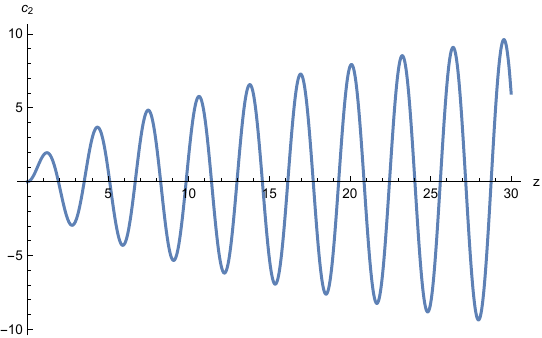}
\includegraphics[width=0.8\textwidth]{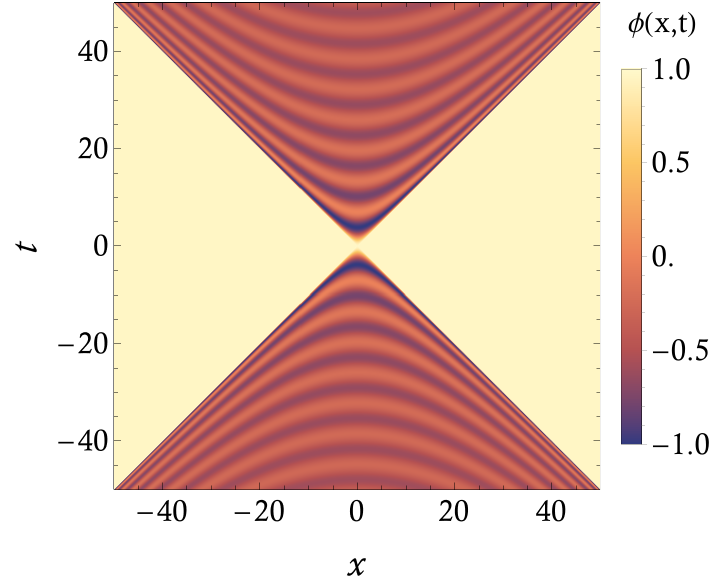}
\caption{Top figures: The dependence of $c_1$ and $c_2$ on $z_0$. Bottom figure: exact antikink kink solution at $v_{in}=1$ 
   }\label{exact}\end{figure}

At large z, the solution has an asymptotic form\cite{DLMFBessel}:
\begin{equation}
    \begin{aligned}
        &J_{0}\left(2 k \sqrt{z}\right)=\sqrt{1/(k\pi \sqrt{z})}\left(\cos\left(2 k \sqrt{z}-%
\tfrac{1}{4}\pi\right)+o\left(1\right)\right)\\
&Y_{0}\left(2 k \sqrt{z}\right)=\sqrt{1/( k\pi \sqrt{z})}\left(\sin\left(2 k \sqrt{z}-%
\tfrac{1}{4}\pi\right)+o\left(1\right)\right).
    \end{aligned}
\end{equation}
 Fig. \ref{exact} shows how $c_1$ and $c_2$ depend on $z_0$. $c_1$ and $c_2$ oscillate with an increasing amplitude. Notice that at $z_0\rightarrow0$, $c_1$ and $c_2$ approach finite values. This limit can be obtained by expanding $c_1$ and $c_2$ in the series of $z$:
 \begin{equation}
     \begin{aligned}
       &  c_1\approx 1+z^2 (-2 \log (z)-2 \gamma +1)+O\left(z^3\right)\\
        & c_2\approx \pi  z^2+O\left(z^3\right),
     \end{aligned}
 \end{equation}
where $\gamma$ is the Euler constant. Therefore, the exact solution at $z_0\rightarrow0$ is
\begin{equation}
    \begin{aligned}
        \phi(x,t)=  \begin{cases}
            J_0\left( \sqrt{t^2-x^2}\right),t^2-x^2\geq 0\\
            \\
            1,t^2-x^2<0.
        \end{cases} 
    \end{aligned}
    \label{exactakkexpress}
\end{equation}
Fig. \ref{exact} (bottom) illustrates the solution at $z_0\rightarrow0$. The boundary of the solution travels at the speed of light and shrinks to a point at $x=0,t=0$. In the bulk, the field oscillates in the vacuum $\phi=1$. Due to its similarity to the AKK collision, we identify this as the AKK collision at $v_{in}=1$.

The exact AKK solution(\ref{exactakkexpress})becomes singular at $x=0$ and $t=0$, as expected. As shown numerically in our model and another compact model\cite{Bazeia:2019tgt}, the collision approaches a more rigid and elastic collision on the compact side.

Another  noteworthy fact is that the Lorentz factor $1/\gamma$ for antikink and kink becomes singular as $v\rightarrow 1$, while in the exact AKK solution (\ref{exactakkexpress}) no such singularity appears. This suggests that at $v\rightarrow1$limit, the superposition of antikink and kink becomes invalid. To obtain a well-defined and stable AKK solution, it is necessary to consider the KAK collision with shock waves. Exploring how to incorporate the shock waves in general kink scattering is an interesting problem that needs further study. 

\subsection{Approximation to the numerical simulation of radiation}
In the bulk,  the exact solution(\ref{analytical sol})and the radiation in KAK for $n\geq 2$ exhibit similar structures. To check their similarity, we fit the numerical simulation result with the analytical solution.

Both kinds of kink collisions mentioned in Section \ref{PDE} oscillate in vacuum $\phi=0$. The first one is the AKK scattering at a speed close to the speed of light. After kinks get reflected, the region between anti-kink and kink oscillates around the vacuum $\phi=0$. The second case is the completely decayed KAK, which appears for $n\geq2$. Except for the finite lifetime oscillon region, all other collisions under the critical velocity result in radiation.

To fit the numerical results, we introduce three parameters, $a,b$, and $t_0$: 
\begin{equation}
    \begin{aligned}
         f(z)=       
            a J_0\left(  \sqrt{(t-t_0)^2-x^2}\right)+ b Y_0\left( \sqrt{(t-t_0)^2-x^2}\right),z\geq z_0.
    \end{aligned}
\end{equation}
$a$ and $b$ are the amplitude of the solution, and  $t_0$ is the position where the field value diverges. To fit these parameters, we use the FindFit function in Mathematica,  minimizing the difference between the first period of the analytical solution and the numerical simulation.

\subsubsection{Vacuum Oscillation in AKK}

\begin{figure}
\centering
\includegraphics[width=0.8\textwidth]{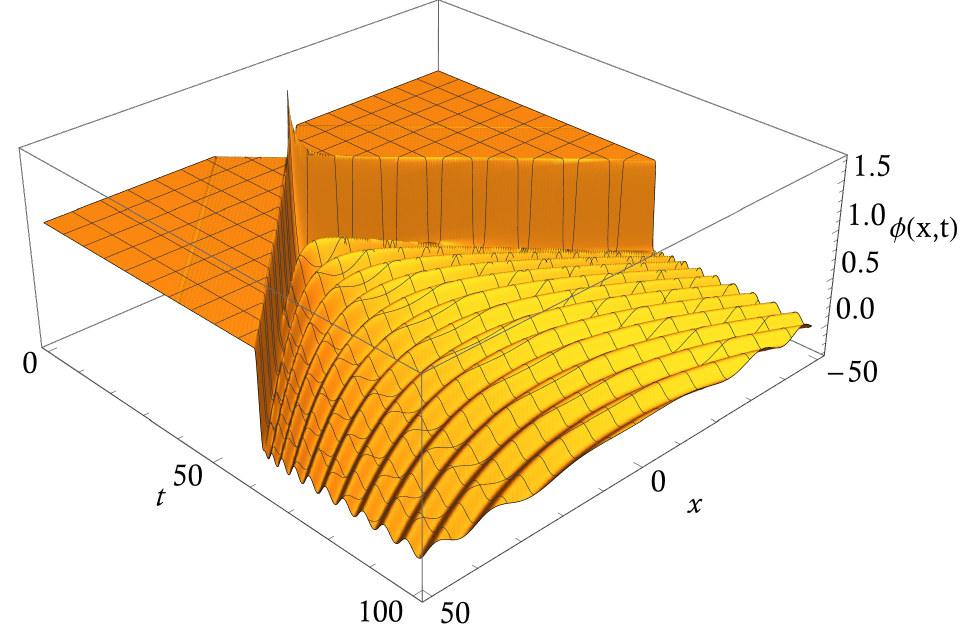}
\includegraphics[width=.45\textwidth]{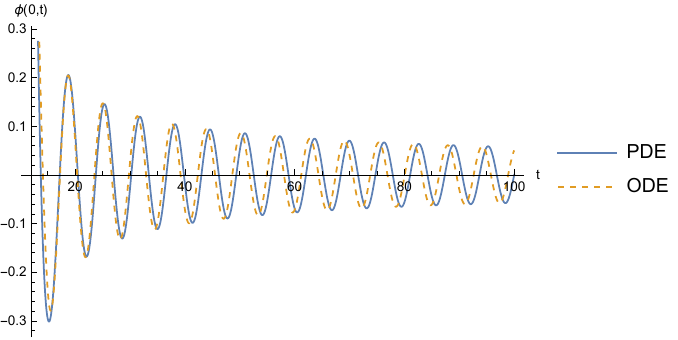}
\quad
\includegraphics[width=0.45\textwidth]{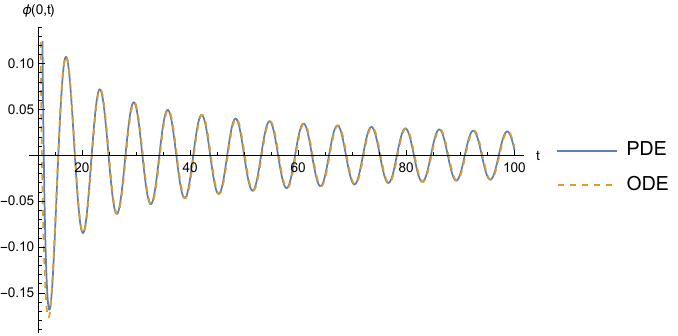}
\caption{Top figure: AKK collision for $n=1$ and $v=0.99$. Bottom figures: The $\phi(0,t)$ plot for numerical result and fitted solution  for $n=1$ and $ n=2$ at  $v_{in}=0.99$. The first period is used for fitting. 
   }\label{2n}\end{figure}

Fig. \ref{2n}(Top Left) shows the AKK vacuum oscillation for the $\phi^6$ model, which is also the $n=1$ case in the quadratic half-compact model. The oscillating region lies between kink and anti-kink after they get reflected. We only fit $\phi(0,t)$ for this configuration. Since the $\phi=1$ boundary of the oscillation moves below the speed of light, the solution close to the boundary has some deviation from the exact solution(\ref{analytical sol}).

Fig. \ref{2n}(bottom) shows the fitting result. In the $\phi^6$ model, the analytical solution doesn't match well with the numerical result, since the higher order terms such as $\phi^6$ and $\phi^4$ in potential can't be neglected. Their agreement becomes much better at $n=2$. This indicates that even at lower n, the vacuum oscillation remains similar to the analytical result. 

Another noteworthy fact is the widespread existence of this kind of vacuum oscillation. For instance, in the $\phi^4$ theory, the kink antikink at $v=0.99$ exhibits a similar structure. This oscillation used to be regarded as the shape mode oscillation\cite{sugiyama}; however, the oscillation becomes radiation-like vacuum oscillation at high velocity. This is not surprising since the region between kink and anti-kink also oscillates in the vacuum $\phi=1$ with a radiation-like boundary. The only difference is that in the $\phi^4$ theory, the nonlinear term has more contribution. 

\subsubsection{Vacuum Oscillation in KAK}

\begin{figure}[htbp]
\centering
\includegraphics[width=.28\textwidth]{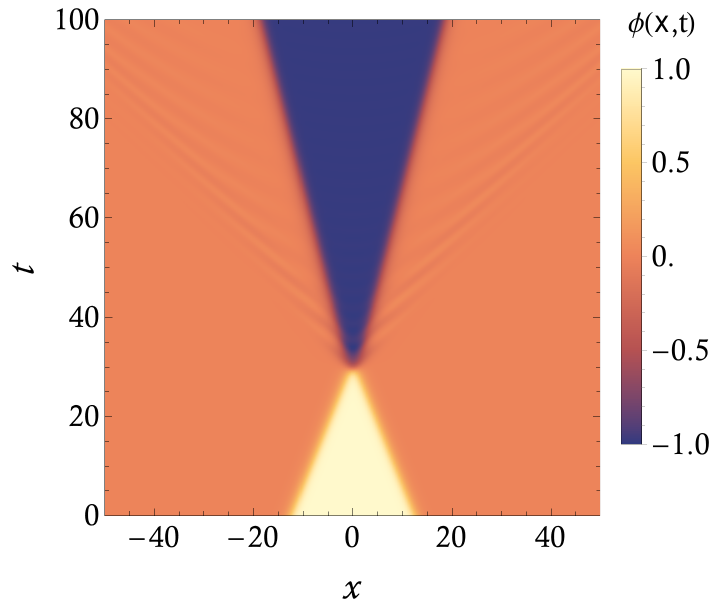}
\qquad
\includegraphics[width=.28\textwidth]{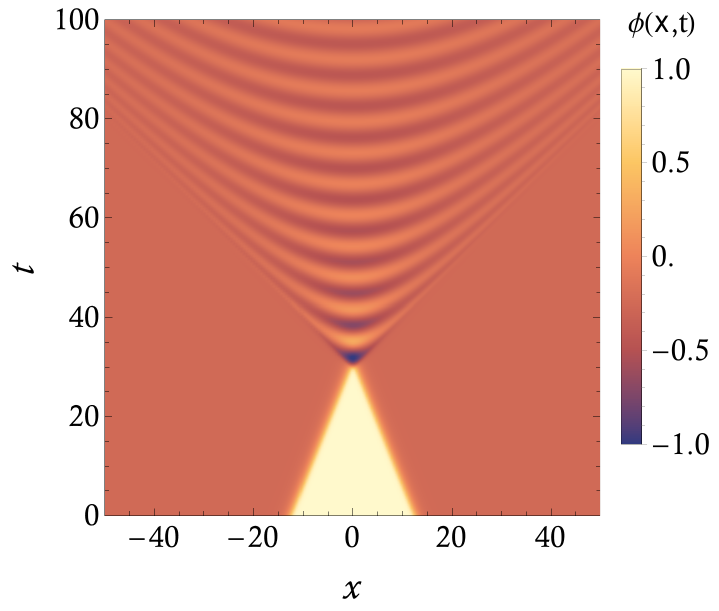}
\qquad
\includegraphics[width=.28\textwidth]{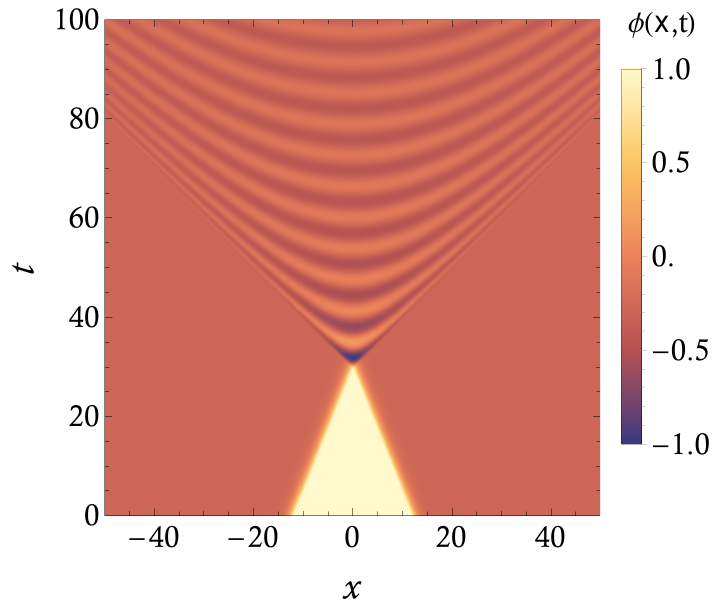}
\qquad
\caption{The evolution of the field in AKK scattering for different n. Left: $n=1,v_{in}=0.4$($\phi^6$ case),  Middle: $n=2,v_{in}=0.4$,  Right: $n=5, v_{in}=0.4$.
\label{3n}}
\end{figure}

For  $n\geq3$ and a part of $n=2$, KAK scattering with incident velocity under the critical velocity decays completely into radiation. Fig. \ref{3n}  shows that this radiation exhibits similar behavior to the vacuum oscillation radiation. Note that in the first figure, the $\phi^6$ case, the radiation exhibits a similar pattern to the second and third figures. This means even the radiation in the kink collision with smaller $n$ could also be related to the analytical solution(\ref{analytical}).

Despite the radiation-like behavior of numerical simulation, the decayed oscillons still have some remaining effects in the completely decayed region. In Fig. \ref{3n}, the amplitude of the first few bounces near $x=0$  is larger than that near the edge, incompatible with the analytical solution. We need to cancel these effects before fitting the two kinds of solutions.

First, since the analytical solution oscillates in a $\phi^2$ potential, the n for the numerical result should also be large enough. We choose $n=5$, which is enough to show their similarity. 

\begin{figure}[htbp]
\centering
\includegraphics[width=.45\textwidth]{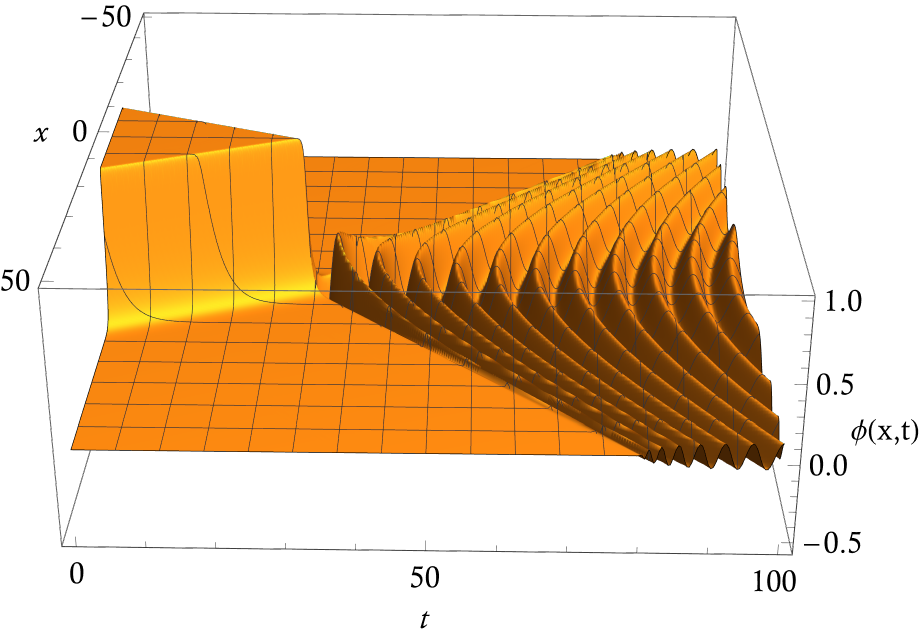}
\qquad
\includegraphics[width=.45\textwidth]{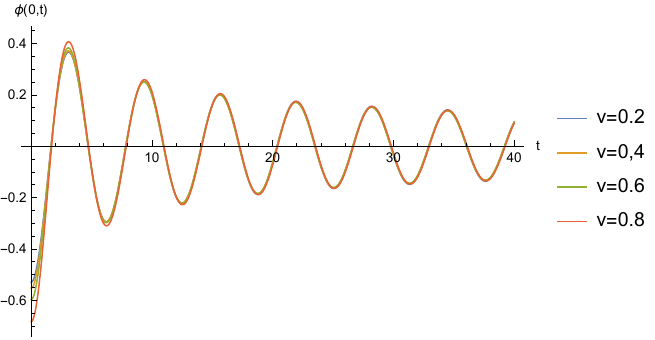}
\qquad
\includegraphics[width=.45\textwidth]{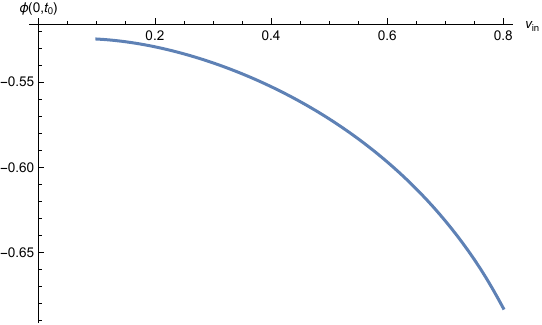}
\qquad
\includegraphics[width=.45\textwidth]{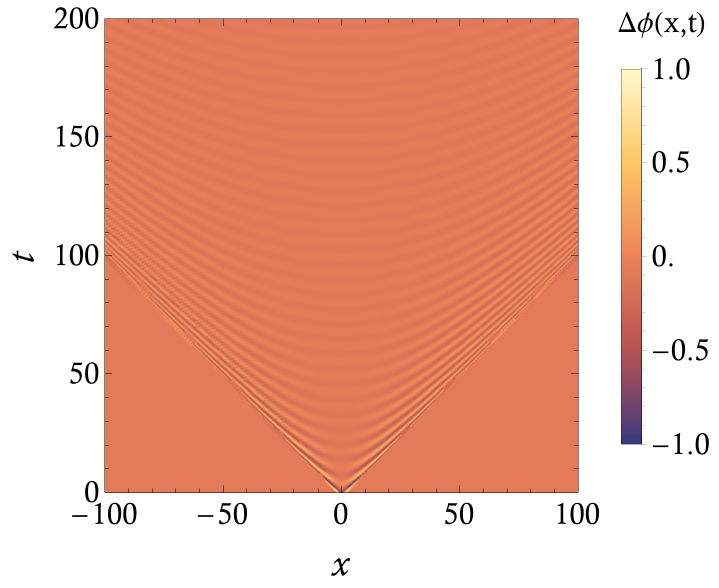}
\qquad

\qquad
\caption{Top Left: The evolution of the field in AKK scattering for $n=3$. Top Right: $\phi(0,t)$ for different $n$. They exhibit similar oscillation at large $z$. Bottom Left: The first peak's amplitude as a function of $v_{in}$. Bottom Right: The difference of the field value at $v_{in}=0.2$ and $v _{in}=0.8$. The first peak of oscillation is moved to the same position.
\label{2v}}
\end{figure}
Next, we consider how the incident velocity affects the remaining oscillon. The bottom right and bottom left plots in Fig. \ref{2v} show that at $n=5$, the amplitude increases with incident velocity. However, at any t, the amplitude of the first few bounces near $x=0$ is always larger than those near larger x with the same $t^2-x^2$. This deviation brings difficulties in fitting, which requires us to fit the configuration after the first few periods. As is illustrated in the top right figure in Fig. \ref{2v}, all the oscillations become the same for large t. The bottom right figure also shows for large $z=\frac{1}{4}   [(t-t_0)^2-x^2 )]$, the four solutions with different $v$ are also similar. Therefore, we will fit the oscillation for larger $z$.

In the AKK case, we check the similarity in the whole configuration. We first check the numerical result's dependence on $z=\frac{t^2-x^2}{4}$, which can be done by finding an orthogonal coordinate. Since the Minkowski spacetime is flat, this is equivalent to finding a diagonal metric. $z$ is chosen to be the same as the Rindler coordinate, with an exchange of t and x. Therefore, we use a Rindler-like coordinate with exchanged $t$ and $x$:

\begin{equation}
    \begin{aligned}
        &x=2\sqrt{z} sinh(y)\\
        &t=2\sqrt{z}cosh(y)+t_0.
    \end{aligned}
    \label{transform}
\end{equation}

 The inverse transformations are:
\begin{equation}
    \begin{aligned}
        &z=\frac{(t-t_0)^2-x^2}{4}\\
        &y=arctanh(\frac{x}{t-t_0}),
    \end{aligned}
\end{equation}
where $t_0$ is the intersection point of the Asymptotes of the radiation. 
The metric in the new coordinate is:
\begin{equation}
    \begin{aligned}
        ds^2=-4zdy^2+\frac{1}{z}dz^2.
    \end{aligned}
\end{equation}
Notice our normalization is slightly different from the standard Rindler coordinate. In this paper, we chose $t_0$ as the local maximum of the second-order derivative. This choice is good enough to show the solution's irrelevance to  y. 
\begin{figure}[htbp]
\centering
\includegraphics[width=.45\textwidth]{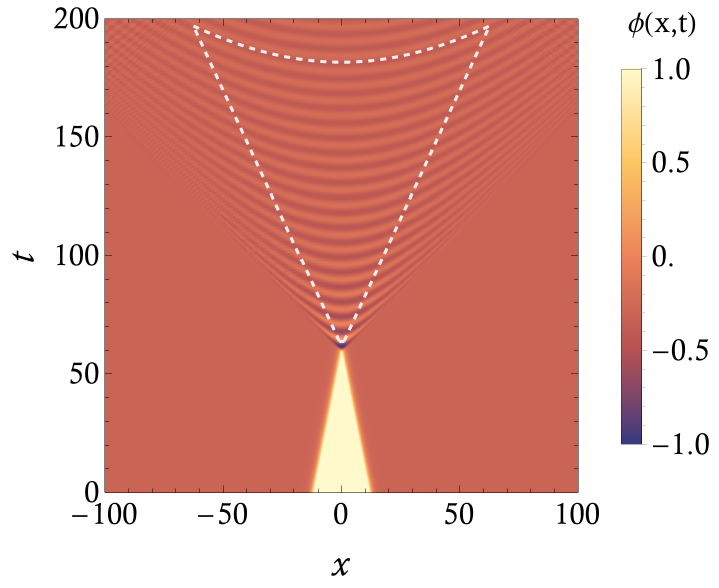}
\qquad
\includegraphics[width=.45\textwidth]{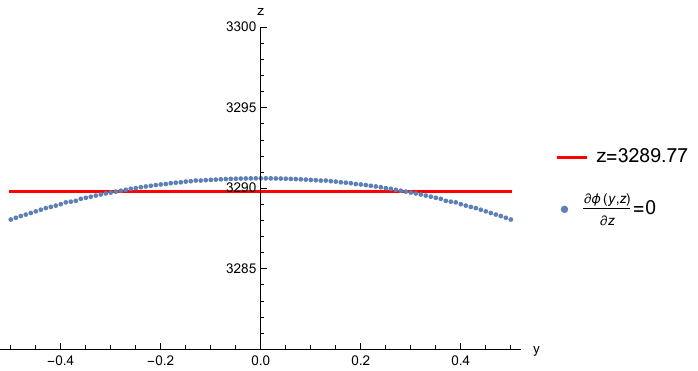}
\qquad
\includegraphics[width=.45\textwidth]{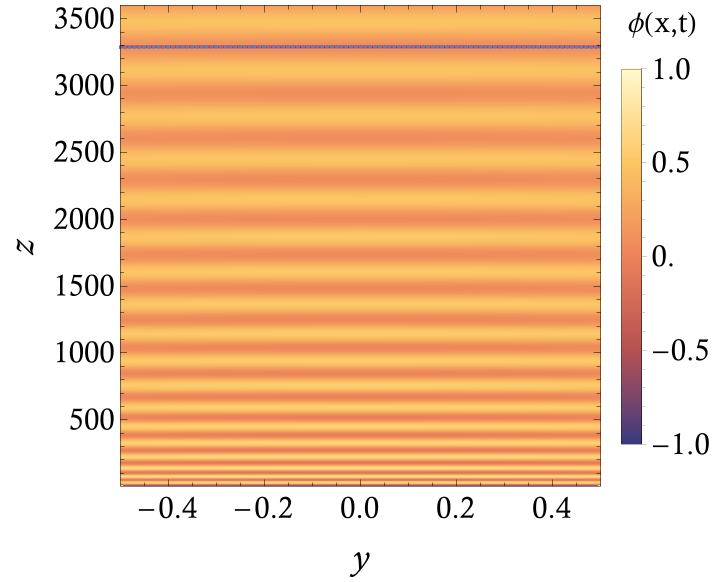}
\qquad
\includegraphics[width=.45\textwidth]{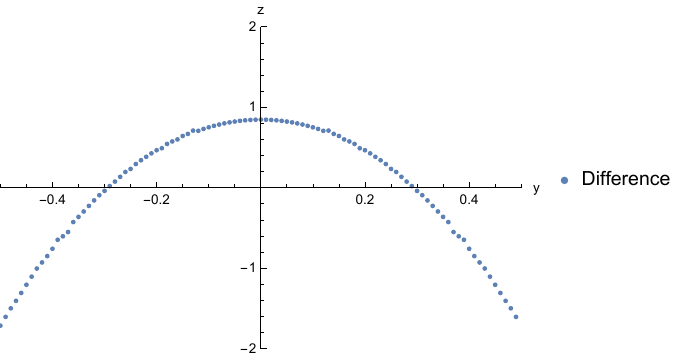}
\caption{Top Left: The AKK scattering at $n=5, v_{in}=0.2$. The dashed region is transformed by the coordinate transformation(\ref{transform})    
Top Right: The fitted line(red) and the curve of peaks in the  $z$-direction.
Bottom Left: The solution after the coordinate transformation.
Bottom Right: The difference between the curve of peaks in the $z$ direction and the fitted line.}\label{coord}
\end{figure}
Fig. \ref{coord}(bottom left) shows the field after coordinate transformation at $n=5,v_{in}=0.2$, $t_0=59.93$. The field value changes slowly with $y$. The top right and bottom right figures of Fig. \ref{coord} also show the local maximum is close to the straight line. Therefore, the solution depends on $y$ weakly.
\begin{figure}[htbp]
    \centering
    \includegraphics[width=.8\textwidth]{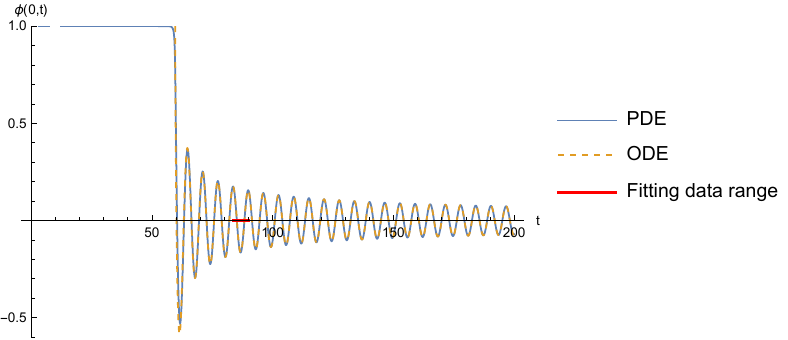}
    \caption{The field $\phi(0,t)$ at $n=5,v_{in}=0.2$, and the fitting ODE solution. The red line is the interval for fitting  }
    \label{4thperiod}  

\end{figure}
Now it is possible to test the similarity between the analytical solution and the numerical simulation in the $t$ direction. Fig. \ref{4thperiod} shows two curves agree well at large $t$.  Therefore, the analytical solution explains the radiation in KAK well.

\begin{figure}[htbp]
    \centering
    \includegraphics[width=.8\textwidth]{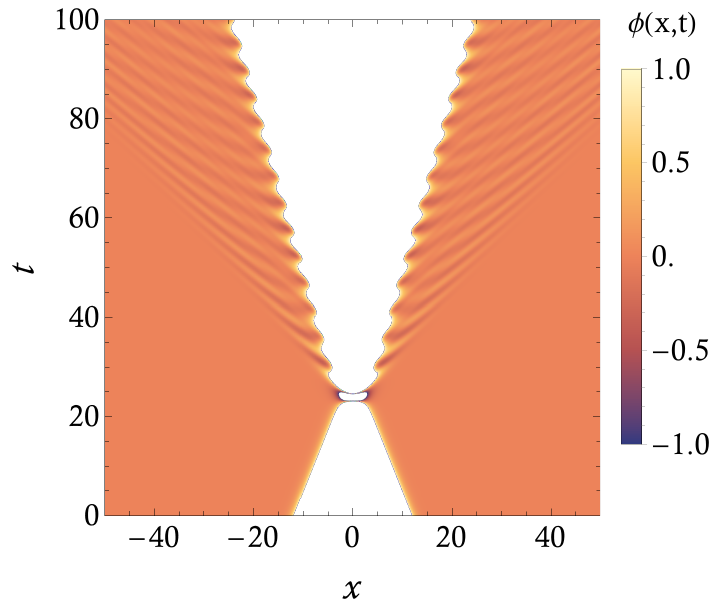}
    \caption{Radiation in $\phi^4$ kink-antikink collision at $v_{in}=0.4$.  }\label{radiationphi4}

\end{figure}

In addition, the radiation in most kink collisions seems to originate from the same mechanism. Since their boundary also moves near the speed of light and stays in a fixed vacuum. Fig. \ref{radiationphi4} illustrates the $\phi^4$ kink-antikink at $v_{in}=0.4$, where the radiation behaves similarly to that in Fig. \ref{3n}. However, the radiation part in the $\phi^6$ and $\phi^4$  theory is hard to be isolated, we leave this as an assumption.

\section{Conclusion}
\label{conclusion}

In this work, we propose a model dependent on the parameter $n$ to investigate vacuum oscillations in kink-antikink collisions, using both numerical and analytical methods. We analyze the fractal structure versus $n$ for both KAK and AKK cases. For KAK scattering, the fractal structure exhibits no resonance bounce phenomena. However, for $n=2$, we find a special velocity interval where the kink and the antikink form an oscillon with a finite lifetime. As the $n$ increases, the kink and antikink become pure radiation at the vacuum $\phi=0$. In the case of AKK scattering, the fractal structures exhibit missing bounce windows with the increase in $n$. The increase in $n$ also restricts the formation of 2-bounce windows, which is likely due to the increase in the delocalized modes.

Notably, we found an exact, radiation-like one-parameter family of  rogue wave solutions \ref{analytical sol} in the $n\rightarrow\infty$ limit.  These solutions agree with the numerical results and further lead to an exact antikink-kink collision solution at $v=1$. This exact solution shows some new behaviors compared to the numerical antikink kink collision at $v\approx 1$, such as the emergence of shock wave structure and the elastic collision behavior. 

 Our result can be generalized to more general radiation in kink collision since the lowest order in potential is often quadratic. For instance, the radiation in the $\phi^4$ and $\phi^6$ theory exhibits similar behaviors. We believe they can also be approximated by exact solutions (\ref{analytical sol}) if the radiation part of the numerical results can be separated. Our work also provides a rare exact kink antikink collision case(\ref{exactakkexpress}). We will study how this solution behaves for large but finite n in future works.

\acknowledgments

We thank Siyu Bian for the helpful discussion and Martin Rocek for pointing out the exact AKK solution (\ref{exactakkexpress}). This research received no specific grant from any funding agency in the public, commercial, or not-for-profit sectors 

The names of the authors are listed alphabetically.
 \bibliographystyle{JHEP}
 \bibliography{biblio.bib}

\end{document}